\def\spacingset#1{\renewcommand{\baselinestretch}{#1}\small\normalsize}

\documentstyle[12pt]{article}
\newcommand{\nc}{\newcommand}
\nc{\rnc}{\renewcommand}
\nc{\nt}{\newtheorem}
\nc{\be}{\begin}
\nc{\erf}[1]{$\ (\ref{#1}) $}
\nc{\rf}[1]{$\ \ref{#1} $}
\nc{\lb}[1]{\mbox {$\label{#1}$} }
\nc{\hr}{\hrulefill}
\nc{\noi}{\noindent}

\nc{\eq}{\begin{equation}}
\nc{\en}{\end{equation}}
\nc{\eqa}{\begin{eqnarray}}
\nc{\ena}{\end{equation}}

\nc{\ra}{\rightarrow}
\nc{\la}{\leftarrow}
\nc{\da}{\downarrow}
\nc{\ua}{\uparrow}
\nc{\Ra}{\Rightarrow}
\nc{\La}{\Leftarrow}
\nc{\Da}{\Downarrow}
\nc{\Ua}{\Uparrow}

\nc{\uda}{\updownarrow}
\nc{\Uda}{\Updownarrow}
\nc{\lra}{\longrightarrow}
\nc{\lla}{\longleftarrow}
\nc{\llra}{\longleftrightarrow}
\nc{\Lra}{\Longrightarrow}
\nc{\Lla}{\Longleftarrow}
\nc{\Llra}{\Longleftrightarrow}

\nc{\mt}{\mapsto}
\nc{\lmt}{\longmapsto}
\nc{\lt}{\leadsto}
\nc{\hla}{\hookleftarrow}
\nc{\hra}{\hookrightarrow}
\nc{\lgl}{\langle}
\nc{\rgl}{\rangle}

\nc{\stla}{\stackrel{d}{\la}}
\nc{\pard}{\partial \da}
\nc{\gdot}{\circle*{0.5}}

\rnc{\baselinestretch}{1.2}      
\nc{\bl}{\vspace{1ex}}           
\rnc{\theequation}{\arabic{section}.\arabic{equation}}  

\nt{thm}{Theorem}[section]
\nt{dfn}[thm]{Definition}
\nt{pro}[thm]{Proposition}
\nt{cor}[thm]{Corollary}
\nt{con}[thm]{Conjecture}
\nt{lem}[thm]{Lemma}

\nc{\Poincare}{\mbox {Poincar$\acute{\rm e}$} }

\nc{\bA}{\mbox {${\bf A}$} }
\nc{\bB}{\mbox {${\bf B}$} }
\nc{\bC}{\mbox {${\bf C}$} }
\nc{\bD}{\mbox {${\bf D}$} }
\nc{\bE}{\mbox {${\bf E}$} }
\nc{\bF}{\mbox {${\bf F}$} }
\nc{\bG}{\mbox {${\bf G}$} }
\nc{\bH}{\mbox {${\bf H}$} }
\nc{\bI}{\mbox {${\bf I}$} }
\nc{\bJ}{\mbox {${\bf J}$} }
\nc{\bK}{\mbox {${\bf K}$} }
\nc{\bL}{\mbox {${\bf L}$} }
\nc{\bM}{\mbox {${\bf M}$} }
\nc{\bN}{\mbox {${\bf N}$} }
\nc{\bO}{\mbox {${\bf O}$} }
\nc{\bP}{\mbox {${\bf P}$} }
\nc{\bQ}{\mbox {${\bf Q}$} }
\nc{\bR}{\mbox {${\bf R}$} }
\nc{\bS}{\mbox {${\bf S}$} }
\nc{\bT}{\mbox {${\bf T}$} }
\nc{\bU}{\mbox {${\bf U}$} }
\nc{\bV}{\mbox {${\bf V}$} }
\nc{\bW}{\mbox {${\bf W}$} }
\nc{\bX}{\mbox {${\bf X}$} }
\nc{\bY}{\mbox {${\bf Y}$} }
\nc{\bZ}{\mbox {${\bf Z}$} }

\nc{\cA}{\mbox {${\cal A}$} }
\nc{\cB}{\mbox {${\cal B}$} }
\nc{\cC}{\mbox {${\cal C}$} }
\nc{\cD}{\mbox {${\cal D}$} }
\nc{\cE}{\mbox {${\cal E}$} }
\nc{\cF}{\mbox {${\cal F}$} }
\nc{\cG}{\mbox {${\cal G}$} }
\nc{\cH}{\mbox {${\cal H}$} }
\nc{\cI}{\mbox {${\cal I}$} }
\nc{\cJ}{\mbox {${\cal J}$} }
\nc{\cK}{\mbox {${\cal K}$} }
\nc{\cL}{\mbox {${\cal L}$} }
\nc{\cM}{\mbox {${\cal M}$} }
\nc{\cN}{\mbox {${\cal N}$} }
\nc{\cO}{\mbox {${\cal O}$} }
\nc{\cP}{\mbox {${\cal P}$} }
\nc{\cQ}{\mbox {${\cal Q}$} }
\nc{\cR}{\mbox {${\cal R}$} }
\nc{\cS}{\mbox {${\cal S}$} }
\nc{\cT}{\mbox {${\cal T}$} }
\nc{\cU}{\mbox {${\cal U}$} }
\nc{\cV}{\mbox {${\cal V}$} }
\nc{\cW}{\mbox {${\cal W}$} }
\nc{\cX}{\mbox {${\cal X}$} }
\nc{\cY}{\mbox {${\cal Y}$} }
\nc{\cZ}{\mbox {${\cal Z}$} }

\nc{\rightcross}{\searrow \hspace{-1 em} \nearrow}
\nc{\leftcross}{\swarrow \hspace{-1 em} \nwarrow}
\nc{\upcross}{\nearrow \hspace{-1 em} \nwarrow}
\nc{\downcross}{\searrow \hspace{-1 em} \swarrow}
\nc{\prop}{| \hspace{-.5 em} \times}
\nc{\wh}{\widehat}
\nc{\wt}{\widetilde}
\nc{\nonum}{\nonumber}
\nc{\nnb}{\nonumber}

\nc{\half}{\mbox {$\frac{1}{2}$} }
\nc{\Cast}{\mbox {$C_{\frac{\infty}{2}+\ast}$} }
\nc{\Casth}{\mbox {$C_{\frac{\infty}{2}+\ast+\frac{1}{2}}$} }
\nc{\Casm}{\mbox {$C_{\frac{\infty}{2}-\ast}$} }
\nc{\Cr}{\mbox {$C_{\frac{\infty}{2}+r}$} }
\nc{\CN}{\mbox {$C_{\frac{\infty}{2}+N}$} }
\nc{\Cn}{\mbox {$C_{\frac{\infty}{2}+n}$} }
\nc{\Cmn}{\mbox {$C_{\frac{\infty}{2}-n}$} }
\nc{\Ci}{\mbox {$C_{\frac{\infty}{2}}$} }
\nc{\Hast}{\mbox {$H_{\frac{\infty}{2}+\ast}$} }
\nc{\Hasth}{\mbox {$H_{\frac{\infty}{2}+\ast+\frac{1}{2}}$} }
\nc{\Hasm}{\mbox {$H_{\frac{\infty}{2}-\ast}$} }
\nc{\Hr}{\mbox {$H_{\frac{\infty}{2}+r}$} }
\nc{\Hn}{\mbox {$H_{\frac{\infty}{2}+n}$} }
\nc{\Hmn}{\mbox {$H_{\frac{\infty}{2}-n}$} }
\nc{\HN}{\mbox {$H_{\frac{\infty}{2}+N}$} }
\nc{\HmN}{\mbox {$H_{\frac{\infty}{2}-N}$} }
\nc{\Hi}{\mbox {$H_{\frac{\infty}{2}}$} }
\nc{\Ogast}{\mbox {$\Omega_{\frac{\infty}{2}+\ast}$} }
\nc{\Ogi}{\mbox {$\Omega_{\frac{\infty}{2}}$} }
\nc{\Wedast}{\bigwedge_{\frac{\infty}{2}+\ast}}

\nc{\Fen}{\mbox {$F_{\xi,\eta}$} }
\nc{\Femn}{\mbox {$F_{\xi,-\eta}$} }
\nc{\Fp}{\mbox {$F_{0,p}$} }
\nc{\Fenp}{\mbox {$F_{\xi',\eta'}$} }
\nc{\Fuv}{\mbox {$F_{\mu,\nu}$} }
\nc{\Fuvp}{\mbox {$F_{\mu',\nu'}$} }
\nc{\cpq}{\mbox {$c_{p,q}$} }
\nc{\Drs}{\mbox {$\Delta_{r,s}$} }
\nc{\spq}{\mbox {$\sqrt{2pq}$} }
\nc{\Mcd}{\mbox {$M(c,\Delta)$} }
\nc{\Lcd}{\mbox {$L(c,\Delta)$} }
\nc{\Wxv}{\mbox {$W_{\chi,\nu}$} }
\nc{\vxv}{\mbox {$v_{\chi,\nu}$} }
\nc{\dd}{\mbox {$\widetilde{D}$} }

\nc{\diff}{\mbox {$\frac{d}{dz}$} }
\nc{\Lder}{\mbox {$L_{-1}$} }
\nc{\bone}{\mbox {${\bf 1}$} }
\nc{\px}{\mbox {${\partial_x}$} }
\nc{\py}{\mbox {${\partial_y}$} }

\begin{document}
\setcounter{equation}{0}
\begin{titlepage}
November 1992
\vspace{.2in}
\begin{center}
\large{{\bf New Perspectives on the BRST-algebraic
Structure of String Theory }}
\end{center}
\vspace{2ex}
\begin{center}
\[
\begin{array}{ccc}
Bong \ H.\ Lian
&  &  Gregg \ J.\ Zuckerman \footnotemark \\
University \ of \ Toronto
& \ \ \ & Yale \ University \\
Department \ of \ Mathematics
& and  &  Department \ of \ Mathematics \\
Toronto, \ Ontario,\ Canada\ M5S\ 1A1
&  &  New \ Haven, \ CT \ 06520, USA  \\
\end{array}
\]
\end{center}
\addtocounter{footnote}{0}\footnotetext{Supported by
NSF Grant DMS-9008459 and DOE Grant DE-FG0292ER25121.
}

\vspace{1ex}
\begin{center}
{\bf Abstract}
\end{center}

Motivated by the descent equation in string theory, we give a new
interpretation
for the action of the symmetry charges on the BRST cohomology in terms of
what we call {\em the Gerstenhaber bracket}. This bracket is compatible
with the graded commutative product in cohomology, and hence gives rise to
a new class of examples of what mathematicians call
a {\em Gerstenhaber algebra}. The latter structure
was first discussed in the context of Hochschild cohomology theory
\cite{Gers1}.
Off-shell in the (chiral) BRST complex, all the identities
of a Gerstenhaber algebra hold up to homotopy.

Applying our theory to the c=1 model, we give a precise conceptual description
of the BRST-Gerstenhaber  algebra of this model. We are  led to
a direct connection between the bracket structure here and
the anti-bracket formalism in BV theory \cite{W2}. We then discuss
the bracket in string backgrounds with both the left and the right movers.
We suggest that the homotopy Lie algebra arising from our
Gerstenhaber  bracket is closely
related to the HLA recently constructed by Witten-Zwiebach.
Finally, we show that our constructions generalize to any topological
conformal field theory.
\end{titlepage}

\section{Introduction}\lb{sec1}

One of the many successes of string theory is to provide a testing ground
for new ideas in physics as well as in mathematics. Often times,
a success story begins with the study of
a certain { \em concrete} model in string
theory (or the cousins thereof). A proper understanding of a special case
leads to generalizations that often go far beyond the original context.
In this paper, we hope to illustrate yet another such episode of the
evolution of string theory.

The enormous success  of the matrix model may be credited for
the recent revival of string theory. This second coming of string theory
marks yet another exciting moment in math/physics. It is perhaps too
soon to give a historical review of this development, for we are still in
the midst of it.
We will however focus on one particular aspect of string theory --
the BRST  structure.

Given a conformal field theory of central charge $c$, one may obtain
a consistent string background by coupling the conformal theory to
both the Liouville theory, with central charge $26-c$, and to the
conformal ghost system, with central charge $-26$.
The background so obtained may then be studied
using CFT techniques,
barring some subtleties coming from the Liouville sector.
The simplest of such backgrounds is the two dimensional string theory,
a.k.a. the c=1 model, in which a single free boson is coupled to the Liouville
field and the ghosts.

It is useful to think of a string background from a slightly more abstract
point
of view. Namely, we can regard a string background as having the form
\be{equation}\lb{1.1}
CFT\otimes ghosts
\end{equation}
The CFT here is a conformal field theory with central charge 26. In the case
of the c=1 model, the CFT may now be viewed as a two dimensional target
spacetime for the string. The symmetry of the target can be exploited
to study the background. For example, this symmetry alone plays a crucial
role in our calculation of the BRST cohomology of the c=1 model
\cite{LZ4}. (Other methods have also been used to  study this problem
\cite{BMP}.) This indicates that some of the detailed structures of the
individual constituents of the CFT in \erf{1.1} may be spared when
one is interested just in the BRST structures of the background.

The second advantage of considering \erf{1.1} abstractly is that
it is easier to pose the question: what are the generic structures
of all string backgrounds?  The abstract setting frees us from some
of the special features and the subtleties of the conformal theory
which one couples to the Liouville field and the ghosts.
Having said that, we will, from now on consider the string backgrounds
of the form \erf{1.1}. Later in the paper we will return to
the c=1 model.

The goal of this paper is to show that for every string background,
the BRST cohomology has,
intrinsically, the algebraic structure known as a Gerstenhaber algebra.
In the case of the c=1 model,
an important {\em realization} of this algebra
is implicit in the work of Witten and Zwiebach\cite{WZ}.
Recently Y. Wu and C. Zhu have reanalysed the same realization
in detail\cite{WuZh}.

In section \ref{sec1.2},
we discuss some basic definitions and results in the theory of
super chiral algebras. In section \ref{sec2}, we present the general
construction of the Gerstenhaber bracket on the BRST cohomology,
and discuss some fundamental properties of the bracket.

Let the dot product, $u\cdot v$, be defined by eqn \erf{2.7}. Let
the bracket product, $\{ u,v\}$, be defined by eqn \erf{2.14}. Let
$|u|$ be the ghost number of $u$. Let $b_0$ be the zero mode
of the anti-ghost field.
   \\
{\bf Theorem 2.2:} {\em On the chiral cohomology $H^*$, we have\\
(a) $u\cdot v=(-1)^{|u||v|}v\cdot u$\\
(b) $(u\cdot v)\cdot t=u\cdot(v\cdot t)$\\
(c) $\{u,v\}=-(-1)^{(|u|-1)(|v|-1)}\{v,u\}$\\
(d)
$(-1)^{(|u|-1)(|t|-1)}\{u,\{v,t\}\}
+(-1)^{(|t|-1)(|v|-1)}\{t,\{u,v\}\}
+(-1)^{(|v|-1)(|u|-1)}\{v,\{t,u\}\}=0$\\
(e) $\{u,v\cdot t\}=\{u,v\}\cdot t +(-1)^{(|u|-1)|v|}v\cdot\{u,t\}$\\
(f) $b_0\{u,v\}=\{b_0u,v\}+(-1)^{|u|-1}\{u,b_0v\}$\\
(g) $\{,\}: H^p\times H^q\lra H^{p+q-1}$.
}
\be{dfn}
An abstract Gerstenhaber algebra $G^*$ is a \bZ-graded vector space equipped
with two bilinear multiplication operations, denoted by $u\cdot v$
and $\{u,v\}$ respectively, and satisfying the following assumptions:\\
(i) If $u$ and $v$ are homogeneous elements of degree $|u|$ and $|v|$
respectively, then $u\cdot v$ is homogeneous of degree $|u|+|v|$ and
$\{u,v\}$ is homogeneous of degree $|u|+|v|-1$.\\
(ii) Identities (a) through (e) from Theorem \ref{thm2.2} above
hold for any triple of homogeneous elements $u,v$ and $t$ in $G^*$.
\end{dfn}

In section \ref{sec3},  we apply our general theory  to the c=1 model.
We give a characterization of the full chiral cohomology algebra.

Let $H^*$ be the chiral cohomology of the c=1 model.
Let $H^*(\pm)$ be subspaces of $H^*$ defined by eqn \erf{3.15}.
  \\

{\bf Theorem 3.2} {\em Let $H^*$ be the chiral cohomology of the c=1 model.
Let $\cA^*$ be the Gerstenhaber algebra
$\bC[x,y,\partial_x,\partial_y]$ (see Appendix B).
Then the following holds:\\
(a) There is an exact sequence of
Gerstenhaber algebras
\[
0\lra H(-)\lra H\lra \cA\lra 0
\]
where $H(-)$ is an ideal in which both products are identically zero.\\
(b) $H^*(+)$ is closed under the dot product, and it is canonically isomorphic
to $\cA^*$, as an associative algebra.\\
(c) Let $H(+)^\sigma$ be the restricted dual of $H(+)$
defined by  the $\bullet$-algebra anti-involution $\sigma$
(section \ref{sec3.5}).
Then the ideal $H(-)$ is isomorphic, as module over $H(+)$, to $H(+)^\sigma$.\\
(d) $H^*(+)$ is {\em not} closed under the bracket product. The
sequence in (a) does not split as an exact sequence of graded Lie algebras
(For more details on the bracket, see section \ref{sec3.6}).\\
(e) The projection map $H\lra\cA$ intertwines $-b_0$ on $H$
and the differential operator
\[
\Delta=\frac{\partial \ }{\partial x}
\frac{\partial \ }{\partial x^*}
+\frac{\partial \ }{\partial y}
\frac{\partial \ }{\partial y^*}
\]
on \cA where $x^*=\px, y^*=\py$.
}

The main purpose of Appendix A is to show that $H^*$
is generated, as a Gerstenhaber algebra, by four generators
$x,y,\partial_x,\partial_y$. Moreover, we describe
a basis of $H(-)$, which is dual and complementary to the
basis of $H(+)$ given in \erf{3.26},
and hence show that $H(-)$ is an ideal with one
generator $\{\px,\py\}$. The pairing between
the two bases is $sl_2$ invariant.
We show, as a consequence, that $H(-)$ is a subalgebra
of $H$ with the zero product. This last assertion can
also be drawn from \cite{WuZh}, where the product
is explicitly computed.
We also describe the action of $b_0,x,y,\px,\py$
on $H(-)$ in terms of the above dual basis.

In Appendix B, we briefly review the classical examples of
Gerstenhaber brackets and algebras. We also attempt to clarify
the relationship between Gerstenhaber brackets and Batalin-Vilkovisky
anti-brackets.

%

{\bf Acknowledgement:} During the course of this work, we have benefitted
from many fruitful conversations with Louis Crane, Igor Frenkel,
James Lepowsky, Bill Massey,
Greg Moore, Ronen Plesser,
Jim Stasheff, Ed Witten, Yongchang Zhu and Barton Zwiebach.

\subsection{Chiral algebras: review}\lb{sec1.2}

In this section, we discuss the definition and some basic
properties of the super chiral algebra (a.k.a vertex operator algebra
in the math literature), which are relevant
to our later discussion. We refer the readers to the original
papers for more extensive discussions on the subject.
The following definition can be found in \cite{FLM}. It is
a refinement of the definition of \cite{BPZ}\cite{Bor}\cite{MS}.
In the context of closed string theory, one should think of
a chiral algebra as a substructure of the full state space of
a conformal string background. In most cases, the full state space
is larger than the chiral algebra itself.

\be{dfn}\lb{def1.1}
A super chiral algebra consists of the following data: a
vector space $V$ graded by
the conformal weight $\Delta$ and the fermion number $|\cdot|$,
two distinguished elements
\bone (the vacuum)  and L (the Virasoro element), and
a one-to-one linear map
\[
\phi\mapsto \phi(z)=\sum\phi_n z^{-n-\Delta_\phi}
\]
where $\phi_n$ is a
linear operator in $V$ of weight $-n$, such that
the product, $\phi(z)\psi$,
is a Laurent series with coefficients in $V$.
The data satisfy the following further conditions:\\
(a) (Cauchy-Jacobi identity) Any two fields $\phi(z),\psi(z) $ satisfy
\be{eqnarray*}
& & Res_w Res_{z-w} (\phi(z-w)\psi)(w) f(z,w)\nonum\\
& & =Res_zRes_w\phi(z)\psi(w) f(z,w)
-(-1)^{|\phi||\psi|}Res_wRes_z\psi(w)\phi(z) f(z,w)
\end{eqnarray*}
where $f$ is any Laurent polynomial in $z,w,z-w$. Note that
the three residues above are taken around the contours with $|w|>|z-w|$,
$|z|>|w|$ and $|w|>|z|$ respectively.\\
(b) The vacuum corresponds to the identity operator;\\
(c) The field $L(z)$ has the OPE
\[
L(z)L(w)\sim \frac{c/2}{(z-w)^4}+\frac{2L(w)}{(z-w)^2}
+\frac{\partial L(w)}{z-w}
\]
and $(L_{-1}\phi)(z)=\partial \phi(z)$. The scalar $c$ is called
the central charge of the chiral algebra.
\end{dfn}
{\bf Note:} Typically physicists denote the operation $Res_w(.)$
by $\int_{C_0} dw(.)$, where $C_0$ is a simple contour surrounding
the point $0$.

There  are many identities that follows from the above definition.
Some may be found in \cite{FLM}\cite{FHL}. Others can be found
in \cite{Zhu}\cite{L2}. Most of the results that we need here
will be derived directly from the Cauchy-Jacobi identity. For example,
if we let $f$ be an arbitrary Laurent polynomial of just  $w$,
and $\phi(z)$ be a current of weight 1, then the Cauchy-Jacobi
identity immediately implies
\be{lem}\lb{lem1.2} The charge $\phi_0=Res_z\phi(z)$ is a derivation
of the product $\psi(w)\chi$, i.e.
\[
\phi_0(\psi(w)\chi)-(-1)^{|\phi||\psi|}\psi(w)(\phi_0\chi)
=(\phi_0\psi)(w)\chi
\]
\end{lem}

Another immediate consequence of the Cauchy-Jacobi identity
is obtained if we set $\psi=\bone$ and $f(z,w)=(z-w)^{-1}g(w)$
where $g(w)$ is an arbitrary Laurent polynomial. In this case,
we get $(\phi_{-\Delta_\phi}\bone)(w)-\phi(w)=0$. Since $\phi\mapsto\phi(w)$
is injective, it follows that
\be{equation}\lb{1.2}
\phi_{-\Delta_\phi}\bone=\phi
\end{equation}
Similarly we have, for
$n>-\Delta_\phi$,
\be{equation}\lb{1.3}
\phi_n\bone=0
\end{equation}

{\bf Notations:} We write $A$ for the state corresponding to the field
$A(z)$, whatever the form of $A(z)$. Thus $\partial\phi$ and
$ce^{\sqrt{2}\phi}$ mean the states corresponding to the fields
$\partial\phi(z)$ and $c(z)e^{\sqrt{2}\phi}(z)$ respectively.

\setcounter{equation}{0}

\section{BRST cohomology and the Gerstenhaber Bracket}\lb{sec2}

In this section, we construct the Gerstenhaber bracket on the chiral
BRST complex \erf{1.1}.
We show that there is a canonical Lie algebra homomorphism
from the space of old physical states
to the ghost number one  BRST cohomology.

\subsection{Setting}\lb{sec2.1}

Consider the chiral algebra of the BRST complex:
\be{equation}\lb{2.1}
C^*=V \otimes \Lambda^*
\end{equation}
where $V$ is the chiral algebra of a conformal field theory with central
charge 26, and $\Lambda^*$ denotes the chiral conformal ghost system.
This space has two standard integral gradings given
by the {\em ghost number} (fermion number) and the {\em conformal weight}.
The subspace of elements with
ghost number $n$ is denoted  $C^n$. We write the ghost number of an
element $a$ as $|a|$.
Every element $a$ of $C^*$  with weight $\Delta$
corresponds uniquely to a field operator
\be{equation}\lb{2.2}
a\mapsto a(z)=\sum a_n z^{-n-\Delta}
\end{equation}
where $a_n$ is a linear operator on $C^*$ and $a_n$  lowers the weight
by $n$.

The chiral algebra $\Lambda^*$ is generated by a pair of fields
$(b(z),c(z))$ of weight (2,-1) and with OPE
\be{equation}\lb{2.2a}
b(z)c(w)\sim\frac{1}{z-w}.
\end{equation}
The stress-energy of this system
is given by
\be{equation}\lb{2.3}
L^\Lambda(z)= 2\partial b(z) \ c(z) + b(z)\partial c(z)
\end{equation}
Here we follow the usual physics convention that whenever two fields
with the same formal variable $z$ are multiplied, the product
actually denotes the normal ordered product.
Since $V$ itself is a chiral algebra, it also has a stress-energy
field which we denote $L^V(z)$.

The (chiral) BRST current is a primary field of weight 1 given by
\be{equation}\lb{2.4}
J(z)=(L^V(z)+\half L^\Lambda(z))c(z) + \frac{3}{2}\partial^2 c(z)
\end{equation}
We denote the BRST charge $J_0$ by $Q$. The (chiral) BRST cohomology
is denoted $H^*$.

The integral gradings on $C^*$ are rather special. The ghost number
 is given by the eigenvalues of the charge $F_0$ of the ghost number
current
\be{equation}
F(z)=c(z)b(z)
\end{equation}
The conformal weight is given by the eigenvalues of $L_0$,
the zero mode of the total stress energy
field
\be{equation}
L(z)=L^V(z)+L^\Lambda(z)
\end{equation}
This field is $Q$-exact because
\be{equation}
[Q,b(z)]=L(z)
\end{equation}
In particular, we have ${[Q,b_0]}=L_0$, which implies that a
BRST invariant state is $Q$-exact unless it has weight zero.

Throughout this paper, $[,]$ will always mean the graded commutator
in some \bZ graded associative algebra \cA. Thus if $u,v$ are
homogeneous elements of \cA, we have
\be{equation}\lb{2.5}
[u,v]=uv-(-1)^{|u||v|}vu
\end{equation}
The vacuum of a chiral algebra will be denoted \bone.

\subsection{The multiplicative structure}\lb{sec2.2}

Since $Q$ is the charge of a current, it acts on the chiral
algebra $C^*$ by derivation (Lemma \ref{lem1.2}), i.e. for any two
homogeneous element $u$ and $v$ in $C^*$,
\be{equation}\lb{2.6}
Q(u(z)v)=(Qu)(z)v +(-1)^{|u|}u(z)(Qv)
\end{equation}
In particular, the bilinear
operation which we call the {\em dot ($\bullet$) product}:
\be{equation}\lb{2.7}
u\cdot v\stackrel{def}{=} Res_z \frac{u(z)v}{z}=
Res_z Res_w \frac{u(z)v(w)\bone}{zw}
\end{equation}
satisfies
\be{equation}\lb{2.6a}
Q(u\cdot v)=(Qu)\cdot v +(-1)^{|u|}u\cdot
(Qv)
\end{equation}
Thus, the dot product induces a product  on the BRST cohomology.
By eqn \erf{1.2}, we have $u\cdot\bone=u=\bone\cdot u$.
Thus $\bone$ is the identity with respect to the dot product.
We claim that eqn\erf{2.7} defines a homotopy graded commutative associative
algebra off-shell, i.e. on the complex $C^*$. This implies in particular
that the product induces a graded commutative associative product on-shell,
i.e. on $H^*$.
The notion of an algebraic identity holding {\em only
} up to homotopy is discussed
in some work of Stasheff \cite{Sta1}\cite{Sta2} (see also \cite{LS}).

 By the Cauchy-Jacobi identity and
eqn\erf{2.7}, we have
\be{eqnarray}\lb{2.8}
u\cdot v&-&(-1)^{|u||v|}v\cdot u \nonum\\
&=& Res_w Res_{z-w}\frac{(u(z-w)v)(w)\bone}{(1+(z-w)/w)w^2} \nonum\\
&=& \sum_{i\geq 0}(-1)^i
Res_w Res_{z-w}\frac{(u(z-w)v)(w)\bone}{ (z-w)^{-i} w^{i+2}} \nonum\\
&=& \sum_{i\geq 0}\frac{(-1)^i}{i+1}
Res_w Res_{z-w}L_{-1}\frac{(u(z-w)v)(w)\bone}{ (z-w)^{-i} w^{i+1}}
\ \ (integration\ by\ parts) \nonum\\
&=& \sum_{i\geq 0}\frac{(-1)^i}{i+1}
Res_w Res_{z-w}
\frac{(Qb_{-1}+b_{-1}Q)(u(z-w)v)(w)\bone}{ (z-w)^{-i} w^{i+1}}\nonum\\
&=& Qm(u,v)+ m(Qu,v)+(-1)^{|u|}m(u,Qv)
\end{eqnarray}
where $m$ is a bilinear operation defined by
\be{equation}\lb{2.9}
m(u,v)= \sum_{i\geq 0}\frac{(-1)^i}{i+1}
Res_w Res_{z-w}\frac{b_{-1}(u(z-w)v)(w)\bone}{ (z-w)^{-i} w^{i+1}}\nonum\\
\end{equation}
Eqn\erf{2.8} says precisely that the dot product is homotopy
graded commutative.

Now consider
\be{eqnarray}\lb{2.10}
(u\cdot v)\cdot t &-& u\cdot (v\cdot t) \nonum\\
&=& Res_w Res_{z-w}\frac{(u(z-w)v)(w)t}{(z-w)w}
-Res_z Res_w \frac{u(z)v(w)t}{zw} \nonum\\
&=& Res_z Res_w \frac{u(z)v(w)t}{(1-w/z)zw}
+(-1)^{|u||v|} Res_w Res_z
\frac{v(w)u(z)t}{(1-z/w)w^2}\nonum\\
& &-Res_z Res_w \frac{u(z)v(w)t}{zw} \nonum\\
&=& \sum_{i>0}
Res_z Res_w \frac{u(z)v(w)t}{z^{i+1}w^{-i+1}}
+(-1)^{|u||v|}\sum_{i\geq0}
Res_w Res_z \frac{v(w)u(z)t}{z^{-i}w^{i+2}}\nonum\\
&=& \sum_{i\geq0}\frac{1}{i+1}
Res_z Res_w \frac{(L_{-1}u)(z)v(w)t}{z^{i+1}w^{-i}}\nonum\\
& &+(-1)^{|u||v|} \sum_{i\geq0}\frac{1}{i+1}
Res_w Res_z \frac{(L_{-1}v)(w)u(z)t}{z^{-i}w^{i+1}}
\ \ (integration\ by\ parts) \nonum\\
&=& \sum_{i\geq0}\frac{1}{i+1}
Res_z Res_w \frac{((Qb_{-1}+b_{-1}Q)u)(z)v(w)t}{z^{i+1}w^{-i}}\nonum\\
& &+(-1)^{|u||v|} \sum_{i\geq0}\frac{1}{i+1}
Res_w Res_z \frac{((Qb_{-1}+b_{-1}Q)v)(w)u(z)t}{z^{-i}w^{i+1}}\nonum\\
&=& Qn(u,v,t)+n(Qu,v,t)+(-1)^{|u|}n(u,Qv,t)
+(-1)^{|u|+|v|}n(u,v,Qt)
\end{eqnarray}
where $n$ is a trilinear operation defined by
\be{eqnarray}\lb{2.11}
n(u,v,t)
&=& \sum_{i\geq0}\frac{1}{i+1}
Res_z Res_w \frac{(b_{-1}u)(z)v(w)t}{z^{i+1}w^{-i}}\nonum\\
& &+(-1)^{|u||v|} \sum_{i\geq0}\frac{1}{i+1}
Res_w Res_z \frac{(b_{-1}v)(w)u(z)t}{ z^{-i} w^{i+1}}
\end{eqnarray}
Eqn\erf{2.10} says that the dot product is homotopy associative.
This proves our claim.

\subsection{The bracket structure}\lb{sec2.3}

Let's motivate the following
construction by something well-known --  the descent
equations. Let $u(z)$ be a BRST invariant field of weight 0. Since
$L_{-1}=[Q,b_{-1}]$, it follows that
\be{equation}\lb{2.12}
\partial u(z)=(L_{-1}u)(z) = (Qb_{-1}u)(z)=[Q,(b_{-1}u)(z)]
\end{equation}
This is an example of a descent equation. Since the left hand side
is a derivative, the coefficient of $z^{-1}$ is zero on both sides.
Thus we have a current $(b_{-1}u)(z)$ whose charge
\be{equation}\lb{2.13}
(b_{-1}u)_0= Res_z (b_{-1}u)(z)
\end{equation}
 is BRST invariant. So if $v$ is a BRST invariant state,
then so is $(b_{-1}u)_0 v$.
In the case of the c=1 model, the {\em formulas} for this operation of
the BRST invariant charges on the BRST invariant states
have been worked out in detail\cite{WZ}.

However, we would like to understand this operation at a
more {\em conceptual} level. The operation
$(b_{-1}u)_0v$ is clearly linear
in both $u$ and $v$. What  is this bilinear operation on
the BRST invariant states? What does it tell us about the cohomology?
What if we extend the operation off-shell?
These are the motivating questions that lead us to study the
bilinear operation. We will introduce the notation
\be{equation}\lb{2.14}
\{ u,v\} = (-1)^{|u|}Res_z (b_{-1}u)(z)v
=(-1)^{|u|}Res_w Res_{z-w} (b(z-w)u)(w)v
\end{equation}
for all $u$, $v$ in $C^*$. Note that this operation decreases
the net ghost number by one, i.e.
$\{,\}: C^p\times C^q\lra C^{p+q-1}$. The sign $(-1)^{|u|}$ is
to make the bracket conform to the convention in \cite{Drin}.
We claim that \\
(i) $Q$ acts by derivation on $\{,\}$;\\
(ii) $\{,\}$ satisfies skew commutativity and the Jacobi identity,
up to homotopy;\\
(iii) this bracket is a biderivation on the dot product, up to homotopy.
Thus the bracket, together with the dot product, defines
a Gerstenhaber algebra structure on the BRST cohomology $H^*$;\\
(iv) $b_0$ acts by derivation on $\{,\}$. Thus
the relative BRST cohomology is closed under the Gerstenhaber
bracket.\\

Let's compute the action of $Q$ on the bracket:
\be{eqnarray}\lb{2.15}
Q\{u,v\}&=&(-1)^{|u|}Res_z {[Q,(b_{-1}u)(z)]}v + (-1)^{|u|}(-1)^{|u|-1}
Res_z (b_{-1}u)(z)Qv \nonum\\
&=&(-1)^{|u|}Res_z (Qb_{-1}u)(z)v - (-1)^{|u|}(-1)^{|u|}
Res_z (b_{-1}u)(z)Qv \nonum\\
&=&(-1)^{|u|-1}Res_z (b_{-1}Qu)(z)v +(-1)^{|u|}Res_z (L_{-1}u)(z)v\nonum\\
& &+ (-1)^{|u|-1}(-1)^{|u|}
Res_z (b_{-1}u)(z)Qv \nonum\\
&=& \{Qu,v\} + (-1)^{|u|-1}\{u,Qv\}
\end{eqnarray}
Note that we have used the fact that $(L_{-1}u)(z)$ is a total
derivative, and hence has zero residue. This proves (i).

Now consider the (graded) skew  commutativity property of $\{,\}$.
{}From eqn\erf{2.14}, it is unclear how this property holds. What we
need is the following elementary but crucial result.
\be{lem}\lb{lem2.1} The following identity holds
\[
(-1)^{|u|}\{u,v\}= b_0(u\cdot v)-(b_0 u)\cdot v -(-1)^{|u|}u\cdot(b_0v)
\]
\end{lem}
A remark about the
identity: the right hand side clearly measures the failure of $b_0$
to be a derivation of the dot product. The same idea first appears
in the "anti-bracket" formalism, but in a seemingly different context.
In \cite{W2}, Witten showed that the Batalin-Vilkovisky equation
can be formulated using a certain fundamental differential operator
$\Delta$ in field space, together with an anti-bracket which
measures the failure of $\Delta$ to be a derivation of an
operator product. The $b_0$ operator here plays the role
of $\Delta$! We will, in Theorem \ref{thm3.2}, make a precise connection
between these two operators in the context of the c=1 model.

The identity above is proved by the following calculation:
\be{eqnarray}\lb{2.16}
& &b_0(u\cdot v)-(b_0 u)\cdot v -(-1)^{|u|}u\cdot(b_0v)\nonum\\
& & =Res_zRes_w\frac{b(z)u(w)v}{z^{-1}w}
-(-1)^{|u|}Res_wRes_z\frac{u(w)b(z)v}{z^{-1}w}
-Res_wRes_{z-w}\frac{(b(z-w)u)(w)v}{(z-w)^{-1}w}\nonum\\
& & =Res_wRes_{z-w}\frac{(b(z-w)u)(w)v}{z^{-1}w}
-Res_wRes_{z-w}\frac{(b(z-w)u)(w)v}{(z-w)^{-1}w}\ \ (Cauchy-Jacobi)\nonum\\
& & =Res_wRes_{z-w}(b(z-w)u)(w)v\nonum\\
& & =(-1)^{|u|}\{u,v\}
\end{eqnarray}

The skew commutativity property of $\{,\}$ now becomes
immediately obvious following Lemma \ref{lem2.1}, at least on-shell.
Let's consider the bracket off-shell.
A simple calculation gives us
\be{eqnarray}\lb{2.17}
\{u,v\}&+&(-1)^{(|u|-1)(|v|-1)}\{v,u\}\nonum\\
&=&(-1)^{|u|-1}(Qm'(u,v)-m'(Qu,v)-(-1)^{|u|}m'(u,Qv))
\end{eqnarray}
where $m'$ is yet another bilinear operation defined by
\be{equation}\lb{2.18}
m'(u,v)=b_0 m(u,v)+ m(b_0u,v) +(-1)^{|u|}m(u,b_0v)
\end{equation}
Thus off-shell, the bracket $\{,\}$ satisfies graded skew commutativity
up to homotopy.

We now consider the graded Jacobi identity. Once again, it is
an easy exercise to show that
\be{eqnarray}\lb{2.19}
& &\{\{u,v\},t\}-\{u,\{v,t\}\}+(-1)^{(|u|-1)(|v|-1)}\{v,\{u,t\}\}\nonum\\
& & =(-1)^{(|v|-1)}
\left(Qn'(u,v,t)+n'(Qu,v,t)+(-1)^{|u|}n'(u,Qv,t)\right.\nonum\\
& &\left. +(-1)^{|u|+|v|}n'(u,v,Qt)\right)
\end{eqnarray}
where $n'$ is a trilinear operation defined by
\be{equation}\lb{2.20}
n'(u,v,t)=Res_z z \ (Res_w b_{-1}(b_{-1}^2u)(w)v)(z)t
\end{equation}
Since $b_{-1}^2=0$, it follows that
$n'$ is identically zero, i.e. the graded Jacobi
identity holds exactly. This completes the proof of (ii).

Consider now the derivation property of the bracket.
\be{eqnarray}\lb{2.21}
\{u,v\cdot t\}&=&(-1)^{|u|}Res_z Res_w \frac{(b_{-1}u)(z) v(w)t}{w} \nonum\\
&=&(-1)^{|u|} Res_w Res_{z-w} \frac{((b_{-1}u)(z-w)v)(w)t}{w}\nonum\\
& &+(-1)^{|u|}(-1)^{(|u|-1)|v|}
Res_wRes_z \frac{v(w)(b_{-1}u)(z)t}{w} \nonum\\
&=& \{u,v\}\cdot t +(-1)^{(|u|-1)|v|} v\cdot\{u,t\}
\end{eqnarray}
This proves that each $\{u,*\}$ is a graded derivation of the dot product.
A similar statement is true for $\{*,t\}$ but only up to homotopy.
More precisely, we have
\be{eqnarray}\lb{2.22}
& &\{u\cdot v,t\}-u\cdot\{v,t\}-(-1)^{(|t|-1)|v|}\{u,t\}\cdot v\nonum\\
& & =(-1)^{|u|+|v|-1}\left( Qn''(u,v,t)-n''(Qu,v,t)\right.\nonum\\
& & \left.-(-1)^{|u|}n''(u,Qv,t)
-(-1)^{|u|+|v|}n''(u,v,Qt)\right)
\end{eqnarray}
where $n''$ is a trilinear operation defined by
\be{equation}\lb{2.23}
n''(u,v,t)=u\cdot m'(v,t)
-m'(u\cdot v,t)
+(-1)^{|t||v|}m'(u,t)\cdot v
\end{equation}
This proves (iii).

Finally, we consider the action of $b_0$ on the bracket.
Using Lemma \ref{lem2.1} and the fact that $b_0^2=0$, we have
\be{equation}\lb{2.24}
b_0\{u,v\}=\{b_0u,v\}
+(-1)^{|u|-1}\{u,b_0v\}
\end{equation}
This proves (iv). Let's summarize all the {\em on-shell} algebraic identities
we have derived.
\be{thm}\lb{thm2.2} On the cohomology $H^*$, we have\\
(a) $u\cdot v=(-1)^{|u||v|}v\cdot u$\\
(b) $(u\cdot v)\cdot t=u\cdot(v\cdot t)$\\
(c) $\{u,v\}=-(-1)^{(|u|-1)(|v|-1)}\{v,u\}$\\
(d)
$(-1)^{(|u|-1)(|t|-1)}\{u,\{v,t\}\}
+(-1)^{(|t|-1)(|v|-1)}\{t,\{u,v\}\}
+(-1)^{(|v|-1)(|u|-1)}\{v,\{t,u\}\}=0$\\
(e) $\{u,v\cdot t\}=\{u,v\}\cdot t +(-1)^{(|u|-1)|v|}v\cdot\{u,t\}$\\
(f) $b_0\{u,v\}=\{b_0u,v\}+(-1)^{|u|-1}\{u,b_0v\}$\\
(g) $\{,\}: H^p\times H^q\lra H^{p+q-1}$.
\end{thm}
Some remarks about these results.
The space $H^0$ is a strict commutative algebra -- a well-known fact.
Note that $H^1$ is closed under
the bracket and hence is an ordinary Lie algebra. Moreover, every
$H^q$ is a module over $H^1$ via the map $(g)$ in the case $p=1$.
If one further restricts to the case $q=0$, one sees that
the Lie algebra $H^1$ acts, by derivations,
on the commutative algebra $H^0$. This crucial fact has been
implicitly used in an effective way to determine the structure
of $H^0$ in the c=1 model \cite{W}.

Note also that $\{,\}$ may be viewed as a super Lie bracket because it
is compatible with the grading ``ghost number - 1''. In particular, we have
$\{H^{r+1},H^{s+1}\}\subset H^{r+s+1}$. The map $u\mapsto \{u,*\}$, which
essentially assigns to each BRST invariant state of weight zero
the corresponding charge $Res_z (b_{-1}u)(z)$,
realizes the adjoint representation of the above super Lie algebra.
This realization is never faithful because $\{\bone,*\}$ is
identically zero.

\subsection{The old physical states}\lb{sec2.4}

We now relate the Lie algebra structure on the old physical states
to that of $H^1$.

Once again let $V$ be a chiral algebra with
central charge $c=26$. The space $P(V)$ of the old physical states
is the subspace of Virasoro primary states of weight 1, modulo the states
of the form $L_n v$, $n<0$. More precisely,
\be{eqnarray}\lb{2.25}
P(V)&=&V[1]^{Vir_+}/N(V)\nonum\\
N(V)&=&(Vir_-V)\cap  V[1]^{Vir_+}
\end{eqnarray}

Let $\phi(z)$ be any primary field of weight 1. This means, in particular,
that the charge $\phi_0$ is $Vir$ invariant:
\be{equation}\lb{2.26}
{[L^V(z),\phi_0]}=0
\end{equation}
It follows that one has a well-defined bilinear operation
\be{equation}\lb{2.27}
\{,\}: V[1]^{Vir_+}\times V[1]^{Vir_+}\lra V[1]^{Vir_+},
\ \ (\phi,\psi)\mapsto -\phi_0\psi
\end{equation}

Let's consider the properties of this operation.
Applying the Cauchy-Jacobi identity, we get
\be{eqnarray}\lb{2.28}
\{\phi,\psi\}
&=&-Res_wRes_z\frac{\phi(w)\psi(z)\bone}{z}
\nonum\\
&=&Res_wRes_{z-w}\frac{(\psi(z-w)\phi)(w)\bone}{(1+(z-w)/w)w}
-Res_zRes_w\frac{\psi(z)\phi(w)\bone}{z}\nonum\\
&=&\sum_{i\geq0}(-1)^iRes_w\frac{(\psi_i\phi)(w)\bone}{w^{i+1}}
\ \ \ (Res_w\phi(w)\bone=0)
\end{eqnarray}
Now note that the $i=0$ term in the last sum is equal to
$-\{\psi,\phi\}$. Thus we have
\be{eqnarray}\lb{2.29}
& &\{\phi,\psi\}
+\{\psi,\phi\}\nnb\\
& &=\sum_{i>0}(-1)^iRes_w\frac{(\psi_i\phi)(w)\bone}{w^{i+1}}\nonum\\
& &=L_{-1}
\sum_{i>0}\frac{(-1)^i}{i}Res_w\frac{(\psi_i\phi)(w)\bone}{w^i}
\ \ \ (integration\ by\ parts)\nnb\\
\end{eqnarray}
This implies that $\{,\}$ is skew-symmetric modulo $N(V)$.
Similarly, it is also easy to
check that the bilinear operation factors through
$N(V)$, and that it satisfies the Lie algebra Jacobi identity
modulo $N(V)$. Thus $\{,\}$ is a Lie bracket on
the space $P(V)$. This Lie algebra structure is already known:
see \cite{FGZ} and references therein.

It is well-known that there are two natural maps
\be{equation}\lb{2.30}
\nu_1,\nu_2 : V[1]^{Vir_+}\lra H^1, H^2
\end{equation}
which send a field $\phi(z)$ to $c(z)\phi(z)$ and $\partial c(z)\ c(z)\phi(z)$
respectively. Let's assume that $V$, as a Virasoro representation,
 has an invariant
bilinear pairing such that the induced
pairing on $H^*$ is non-degenerate. Then we have a pairing
preserving map
\be{equation}\lb{2.31}
\nu:V[1]^{Vir_+}\lra H^1\oplus H^2,\ \ \
\phi\mapsto\frac{1}{\sqrt{2}}(\nu_1\phi\oplus\nu_2\phi)
\end{equation}
Since $N(V)$ lies in the kernel of the pairing on $V$, under the map
$\nu$, the space $N(V)$ must be sent to zero. Thus the map $\nu$ and
hence $\nu_1$ factors through $N(V)$. So we have the map
\be{equation}\lb{2.32}
\nu_1: P(V)\lra H^1
\end{equation}

We claim that this is a Lie algebra homomorphism. By definition
\erf{2.14}, we have
\be{eqnarray}\lb{2.33}
\{\nu_1\phi,\nu_1\psi\}&=& -Res_z(b_{-1}\nu_1\phi)(z)\nu_1\psi\nonum\\
&=& -Res_z(b_{-1}c_1\phi)(z)c_1\psi\nonum\\
&=& -Res_z c_1\phi(z)\psi\nonum\\
&=&\nu_1\{\phi,\psi\}
\end{eqnarray}
which proves the claim.
In case the map $\nu_1$ is injective (which happens whenever
$N(V)$ coincides with the kernel of the pairing on $V$), such
as in the case of the 26 dimensional bosonic string theory,
we have a concrete realization of the Lie subalgebra $P(V)$ of
$H^1$.

\setcounter{equation}{0}

\section{The c=1 model}\lb{sec3}

In this section, we will apply the machinery of the last section
to obtain a characterization of the cohomology algebra of the
c=1 model.

Let's first review the operator formalism of the
c=1 model. The model may be described
as a theory in which a single free boson $X$ is coupled to
the Liouville field $\phi$ and the conformal ghosts $(b,c)$.
As we did in \cite{LZ5}, we consider the case in which the cosmological
constant is zero. For now, we restrict ourselves to the
holomorphic part of the theory.

\subsection{What's known}\lb{sec3.1}

The free boson sector can be simply described by
the field $X(z)$ with the OPE
\be{equation}\lb{3.1}
X(z)X(w)\sim -ln(z-w)
\end{equation}
The operators in this sector are linear combinations of
\be{equation}\lb{3.2}
P(X) e^{ipX}(z)
\end{equation}
where  the $P(X)$ are  polynomials in the derivatives of
$X(z)$, and $p$ is the {\em momentum}
of the operator \erf{3.2}.
The corresponding state space is a direct sum of highest weight
representations $F(p)$ of the Heisenberg algebra:
\be{equation}\lb{3.3}
{[\alpha_n,\alpha_m]}=n\delta_{n+m,0}
\end{equation}
where $i\partial X(z)=\sum\alpha_n z^{-n-1}$.
The stress-energy field is
\be{equation}\lb{3.4}
L^X(z)=-\half (\partial X(z))^2
\end{equation}

The Liouville sector can be described in a similar way -- with $X$
replaced by $\phi$ everywhere -- except
that the stress-energy of this sector is given by
\be{equation}\lb{3.5}
L^\phi(z)=-\half(\partial\phi(z))^2+\sqrt{2}\partial^2\phi(z)
\end{equation}
We write $i\partial\phi(z)=\sum j_n z^{-n-1}$, where
the $j_n$ represent a Heisenberg algebra just like the $\alpha_n$ do.
Anticipating that we eventually will deal only with pure imaginary
Liouville momenta, we denote the Liouville state space by $F(-i\alpha)$.
The $\alpha$ will be restricted to real later. But for now, $\alpha$ is
arbitrary.

The ghost sector $\Lambda^*$
 has already been described, so we won't
repeat it here.
Now the BRST operator $Q$ acts on
the spaces
\be{eqnarray}\lb{3.6}
& &C^*(p,\alpha)=F(p)\otimes F(-i\alpha)\otimes\Lambda^*
\nonum\\
& & Q: C^n(p,\alpha)\lra C^{n+1}(p,\alpha)
\end{eqnarray}
Thus we have one cochain complex for every pair of momenta $p,\alpha$.
Later we will restrict the momenta to lie in a certain  two dimensional
even lattice. This restriction will land us back in
the framework we considered in section \ref{sec2}.

The cohomology of the complexes $C^*(p,\alpha)$ was first studied
by us, in connection with the $c<1$ models\cite{LZ3}\cite{L1}.
This cohomology problem has a number of other interesting applications.
In \cite{LZ4}, we applied the results to the c=1 model.
(Subsequently, other methods were also used to analyze the cohomology
of the $c\leq 1$ models
\cite{BMP}\cite{MMS}.) Our cohomology results
have also led to some later work \cite{FeFr?} which gains
new insights into the  structure  theory
of the Feigin-Fuchs representations of the Virasoro algebra.

Our results on the c=1 model have recently been given an interesting
physical interpretation \cite{W} in connection with the matrix model.
The object that plays a key role in this interpretation is what's
called ``the ground ring.'' This is the subalgebra $H^0$
of the associative algebra $H^*$. The main tool that was used
to determine the structure of the ground ring was
the action coming from the tachyon fields. (Actually,
this is secretly a part of the action of $H^1$
on $H^0$ via the Gerstenhaber bracket, as we shall see.)
The symmetry of the c=1 model has been better understood by means of
the ground ring.
More recently, the role of the symmetry has been further
clarified in the context of closed string theory \cite{WZ}.

Let's now return to the cohomology problem (see \cite{LZ4}\cite{BMP}
and references
therein).
\be{thm}\lb{thm3.1}
The cohomology $H^*(p,\alpha)$ of the complex $C^*(p,\alpha)$
is given as follows:\\
(a) $H^*(p,\alpha)$ is nontrivial iff either (i) $(p,\alpha)$ lies
in one of the two {\em complex} lines defined by
$(\alpha-\sqrt{2}+p)(\alpha-\sqrt{2}-p)=0$;
or (ii) $(p,\alpha)$ lies in the intersection of the even lattice \cL
and the past-future cone \cC in the {\em real} $p\alpha$-plane.
Here \cL and \cC are defined by (Figure 1):
\be{eqnarray*}
\cL&=&\{(p,\alpha)\ |\ p-\alpha\in\sqrt{2}\bZ, \ p,\alpha\in \bZ/\sqrt{2}\ \}\\
\cC&=&\{(p,\alpha)\in\bR^2\ | either\ (\alpha-\sqrt{2}+p)\ and\
(\alpha-\sqrt{2}-p)\ \\
& &both\ greater\ than\ zero(future)\ ,\ or\ both\ less\ than\ zero(past)\ \}
\end{eqnarray*}
(b) The dimensions of the $H^*(p,\alpha)$ in case (i) are
\[
dim\ H^n(p,\alpha)=\left\{ \be{array}{ll}
  1 & \mbox{if $n=1,2$}\\
  0 & \mbox{otherwise}
\end{array}
\right.
\]
In case (ii) (hence $(p,\alpha)$ lies in \cL),
we have $H^n(p,\alpha)=0$ unless $0\leq n\leq 3$, and
\be{eqnarray*}
dim\ H^0(p,\alpha)&=&\left\{ \be{array}{ll}
  1 & \mbox{if $(p,\alpha)$ in the past cone }\\
  0 & \mbox{if $(p,\alpha)$ in the future cone }\\
\end{array}
\right.\\
dim\ H^1(p,\alpha)&=&\left\{ \be{array}{ll}
  2 & \mbox{if $(p,\alpha)$ in the past cone }\\
  1 & \mbox{if $(p,\alpha)$ in the future cone }\\
\end{array}
\right.\\
dim\ H^2(p,\alpha)&=&\left\{ \be{array}{ll}
  1 & \mbox{if $(p,\alpha)$ in the past cone }\\
  2 & \mbox{if $(p,\alpha)$ in the future cone }\\
\end{array}
\right.\\
dim\ H^3(p,\alpha)&=&\left\{ \be{array}{ll}
  0 & \mbox{if $(p,\alpha)$ in the past cone }\\
  1 & \mbox{if $(p,\alpha)$ in the future cone }\\
\end{array}
\right.
\end{eqnarray*}
\end{thm}
The states in case (i) are basically the
tachyon states (and their duals).

\subsection{Witten's ground ring}\lb{sec3.2}

To return to the framework of section \ref{sec2}, we will
restrict the momentum values $(p,\alpha)$ to the lattice \cL.
It is known that the space
\be{equation}\lb{3.7}
V= \oplus_{(p,\alpha)\in\cL} F(p)\otimes F(-i\alpha),
\end{equation}
together with the grading coming from the spectrum of
$L_0=L^X_0+L^\phi_0$, forms a chiral algebra. Thus
the super chiral algebra
\be{equation}\lb{3.8}
C^*=V\otimes\Lambda^*
\end{equation}
is example of the situation we considered in section \ref{sec2}.
In particular, the BRST cohomology $H^*$ of the complex $C^*$
has all the structures stated in Theorem \ref{thm2.2}.
In particular, $H^0$ is a {\em commutative} algebra.

Witten proves that $H^0$ is a polynomial algebra with
two generators, which are represented by
\be{eqnarray}\lb{3.9}
\cO_{1/2,1/2}&=&\left(cb+\frac{i}{\sqrt{2}}(\partial X-i\partial\phi)\right)
e^{(iX-\phi)/\sqrt{2}}\nonum\\
\cO_{1/2,-1/2}&=&\left(cb-\frac{i}{\sqrt{2}}(\partial X+i\partial\phi)\right)
e^{(-iX-\phi)/\sqrt{2}}\nonum\\
\end{eqnarray}
Let's briefly recap his argument.
First it is shown that there are
two special derivations, which we denote $\delta_\pm$,
acting on the ground ring. In particular, they act on
$\cO_{1/2,\pm 1/2}$ by
\be{eqnarray}\lb{3.10}
\delta_\pm \cO_{1/2,\pm 1/2}&=&1\nonum\\
\delta_\pm \cO_{1/2,\mp 1/2}&=&0
\end{eqnarray}
This immediately implies that all the monomials generated by
$\cO_{1/2,\pm 1/2}$ are necessarily nonzero. Moreover, it is
easy to see that all such monomials have distinct momenta.
Specifically, the monomials
$\cO_{1/2,1/2}^n
\cdot\cO_{1/2,-1/2}^m$ have momenta
$(p,\alpha)=(\frac{n-m}{\sqrt{2}},\frac{-n-m}{\sqrt{2}})$.
Thus they must be linearly independent.
Now by Theorem \ref{thm3.1} part (b), case (ii),
we see that the momenta are multiplicity free in ghost number zero.
This proves that the above monomials exhaust all of $H^0$.
To summarize, we have an isomorphism of commutative algebras
\be{equation}\lb{3.11}
\psi : H^0\lra \bC[x,y],\ \ \ \cO_{1/2,1/2},\cO_{1/2,-1/2}\mapsto x,y
\end{equation}

In a more recent paper of Witten and Zwiebach \cite{WZ}, the structure
of the cohomology algebra $H^*$ has become better understood.
For example, it has been indicated that $H^*$ contains a subalgebra
which is isomorphic to the polynomial super algebra
\be{equation}\lb{3.11a}
\cA=\bC{[x,y,\partial_x,\partial_y]}
\end{equation}
 where $x,y$ are bosonic and
$\partial_x,\partial_y$ are fermionic.
What can we say about the dot and bracket products on the full
cohomology space $H^*$?
This is the subject of the next discussion.
It is also one of the main applications
of our theory in section \ref{sec2}.

\subsection{Extending the map $\psi$}\lb{sec3.3}

The polynomial super algebra \cA above may be thought of as
a space of polyvector fields on the $xy$-plane.
It was known to Schouten \cite{Sch} that the space
of polyvector fields admits a bracket operation $\{,\}$ which extends
the Lie bracket on 1-vector fields, and extends
the action of the 1-vector fields
on the algebra of functions (0-vector fields).
In the case of the graded algebra \cA, this bracket is uniquely characterized
by the identities (see also Appendix B):\\
(i) $|x|=|y|=0$; $|\partial_x|=|\partial_y|=1$\\
(ii) $\{x,y\}=\{\partial_x,y\}=\{\partial_y,x\}=0$\\
(iii) $\{\partial_x,x\}=\{\partial_y,y\}=1 $\\
(iv) $\{u,v\}=-(-1)^{(|u|-1)(|v|-1)}\{v,u\}$\\
(v) $\{u,v\cdot t\}=\{u,v\}\cdot t+(-1)^{(|u|-1)|v|}v\cdot \{u,t\}$\\
where $u,v,t$ are any polyvector fields.
Thus, \cA becomes a Gerstenhaber algebra. We denote by $\cA^p$ the
subspace of $p$-vector fields.

Note that every 1-vector field is uniquely determined by a
derivation on the algebra of 0-vector fields $\cA^0$.
Similarly, every 2-vector field $f\cdot\partial_x\cdot\partial_y$
is determined by the operation
\be{equation}\lb{3.12}
\{f\cdot\partial_x\cdot\partial_y,*\}:
\cA^0\lra \cA^1
\end{equation}
Recall the homomorphism $\psi$ (see \erf{3.11}).
The crucial things to notice are that
$\psi$ is an {\em isomorphism},
and that {\em both}
$H^*$ and $\cA^*$ have a bracket structure.
We can therefore extend $\psi$ in such a way that
makes the two brackets compatible,    as follows.

Given a ghost number 1 class $u$ in $H^1$, we let $\psi u$ be
the 1-vector field satisfying
\be{equation}\lb{3.13}
\{\psi u,f\}=\psi\{u,\psi^{-1}f\}, \ \  f\ is\ any\ polynomial\ function
\end{equation}
Note that the bracket on the left hand side is for $\cA^*$,
while the one on the right is for $H^*$.
Because $\{u,\psi^{-1}f\}$ is in $H^0$, the right hand side
is well-defined.
The map has now been extended to $\psi:H^1\lra \cA^1$.

Similarly, given a ghost number 2 class $v$,
we let $\psi v$ be the 2-vector field satisfying
\be{equation}\lb{3.14}
\{\psi v,f\}=\psi\{v,\psi^{-1}f\}
\end{equation}
Since $\psi$ is well-defined on $H^1$, the right hand side
of eqn\erf{3.14} makes sense.
Now since there is no $\cA^3$, we set $\psi H^3=0$.
Therefore, $\psi$ is now defined on all of $H^*$.
Note that $\psi$ obviously preserves the ghost number.

Since $\cA^0$ is generated by $x,y$, the value of
$\psi u$ is determined by  the cases $f=x,y$ in eqn\erf{3.13}.
The same is true for $\psi v$ in
eqn\erf{3.14}.
We now use this observation together
with Theorem \ref{thm3.1} to understand the kernel and the
cokernel of $\psi$.

\subsection{The structure of $\psi$}\lb{sec3.4}

Let's first introduce some notations. For every real number
$\alpha_0$, we denote by $H^n(\alpha\geq\alpha_0)$ the sum of
all $H^n(p,\alpha)$ with $(p,\alpha)$ in the lattice \cL
and $\alpha\geq\alpha_0$. Similarly
for $H^n(\alpha<\alpha_0)$ etc.
We also write
\be{eqnarray}\lb{3.15}
H(+)&=&H^0(\alpha\leq0)\oplus H^1(\alpha\leq\sqrt{2}/2)
\oplus H^2(\alpha\leq\sqrt{2})\nonum\\
H(-)&=&H^3(\alpha\geq2\sqrt{2})\oplus H^2(\alpha\geq3\sqrt{2}/2)
\oplus H^1(\alpha\geq\sqrt{2})
\end{eqnarray}
The space $H(+)$ ($H(-)$) is what Witten-Zwiebach called the
plus (minus) states \cite{WZ}. Note that by Theorem \ref{thm3.1},
(see also Figure 2) $H^0(\alpha>0)$ and $H^3(\alpha<2\sqrt{2})$ are both zero.
Thus $H(\pm)$ are actually two {\em complementary} subspaces of $H$.

We claim that $\psi$ is a Gerstenhaber algebra homomorphism, and that
\be{eqnarray}\lb{3.16}
ker\ \psi&=&H(-)\nonum\\
im\ \psi&=&\cA\cong H(+)\ \ \ as\ \bullet- algebras
\end{eqnarray}
We prove these results in stages:\\
(i) $\psi H^1(\alpha\geq\sqrt{2})=\psi H^2(\alpha\geq3\sqrt{2}/2)=0$;\\
(ii) $\psi(u\cdot v)=\psi(u)\cdot\psi(v)$;\\
(iii) $\psi\{u,v\}=\{\psi u,\psi v\}$;\\
(iv) $\psi$ is onto, and when restricted to $H(+)$,
is a $\bullet$-algebra isomorphism.\\
Because $\psi H^3=0$ by definition,
part (i) proves $H(-)\subset ker\ \psi$.
Since $H(+)$ is a complementary subspace of $H(-)$
in $H$, (iv) proves eqns \erf{3.16}.
Parts (ii) and (iii) prove that $\psi$ is a Gerstenhaber
algebra homomorphism.

Recall that the states $\cO_{1/2,\pm 1/2}$ have Liouville
momenta $\alpha=-\sqrt{2}/2$.
By definition \erf{3.13} and by momentum conservation, we have
\be{equation}\lb{3.17}
\{\psi H^1(\alpha\geq\sqrt{2}), x\ or\ y\}
=\psi\{ H^1(\alpha\geq\sqrt{2}),\cO_{1/2,\pm 1/2}\}
\subset\psi H^0(\alpha\geq\sqrt{2}/2)
\end{equation}
The last space is zero as noted earlier.
The condition \erf{3.17} implies that
$\psi H^1(\alpha\geq\sqrt{2})=0$.
Similarly, we have
\be{eqnarray}\lb{3.18}
\{\psi H^2(\alpha\geq3\sqrt{2}/2), x\ or\ y\}
&=&\psi\{ H^2(\alpha\geq3\sqrt{2}/2),\cO_{1/2,\pm 1/2}\}\nonum\\
&\subset&\psi H^1(\alpha\geq\sqrt{2})\nonum\\
&=&0
\end{eqnarray}
This implies that $\psi H^2(\alpha\geq3\sqrt{2}/2)=0$, and hence
completes the proof of part (i).

To prove part (ii), we need to check that it holds for
$u\in H^r$, $v\in H^s$ for any $0\leq r,s\leq3$. But note that (ii)
trivially holds whenever $r+s\geq3$, for then $\cA^{r+s}=0$
and $\psi H^{r+s}=0$.
It holds for $r=0,s=0$ because $\psi$ restricted to $H^0$ is
a $\bullet$-algebra isomorphism. Let consider the first nontrivial
case $u\in H^0$, $v\in H^1$.
\be{eqnarray}\lb{3.19}
\{\psi(u\cdot v),f\}&=& \psi\{u\cdot v,\psi^{-1}f\}\nonum\\
&=& \psi(u\cdot\{v,\psi^{-1}f\})\ \ \ (\{H^0, H^0\}=0)\nonum\\
&=& \psi(u)\cdot\psi\{v,\psi^{-1}f\}\ \ \ (by\ the\ case\ r=0,s=0) \nonum\\
&=& \psi(u)\cdot\{\psi(v),f\}\ \ \ (by\ definition\ of\ \psi\ on\
 H^1) \nonum\\
&=& \{\psi(u)\cdot\psi(v),f\}\ \ \ (\{\cA^0, \cA^0\}=0)
\end{eqnarray}
This implies that $\psi(u\cdot v)=\psi(u)\cdot\psi(v)$.
Now all the remaining cases can be handled the same way.
This proves part (ii).

Since $\{H^r,H^s\}\subset H^{r+s-1}$, part (iii)
holds trivially in the case $r+s\geq4$ or the case
$r+s\leq0$. In the case $r=0,s=1$, part (iii) follows directly
from the definition of $\psi$ in eqn \erf{3.13}.
In the case $r=0,s=3$, it holds for the following reason.
By momentum conservation, $\{H^3,\cO_{1/2,\pm 1/2}\}\subset
H^2(\alpha\geq3\sqrt{2}/2)$. But the right hand
side gets sent to zero by part (i). This implies that
$\psi\{H^3,\cO_{1/2,\pm 1/2}\}=0$, hence
$\psi\{H^3,H^0\}=0$. Since $\psi H^3=0$,
it follows that both sides of part (iii) are zero.

There are three remaining nontrivial cases to check:
$(r,s)=(1,1),(0,2),(1,2)$. We will do the first one,
while the rest are similar. Let $u,v\in H^1$. Then $\{u,v\}\in H^1$.
Thus we have
\be{eqnarray}\lb{3.20}
\{\psi\{u,v\},f\}&=& \psi\{\{u,v\},\psi^{-1}f\}\nonum\\
&=&\psi\{u,\{v,\psi^{-1}f\}\}-\psi\{v,\{u,\psi^{-1}f\}\}\ \ \
(by\ Jacobi\ id\ for\ H^*)\nonum\\
&=&\{\psi u,\psi\{v,\psi^{-1}f\}
-\{\psi v,\psi\{u,\psi^{-1}f\}\ \ \ (by\ the\ case\ r=0,s=1)\nonum\\
&=&\{\psi u,\{\psi v,f\}
-\{\psi v,\{\psi u,f\}\ \ \ (by\ definition\ of\ \psi\ on\ H^1 )\nonum\\
&=&\{\{\psi u,\psi v\},f\}\ \ \ (by\ Jocobi\ id\ for\ \cA^*)
\end{eqnarray}
This completes the proof of part (iii).

To prove the first half of part (iv), it is enough to show that
the fermionic generators $\partial_x,\partial_y$ of \cA are
in the image of $\psi$.
This has basically been done already in \cite{W}\cite{WZ}.
Consider the following BRST invariant states:
\be{equation}\lb{3.21}
Y^+_{1/2,\pm1/2}=-ce^{(\pm iX+\phi)/\sqrt{2}}
\end{equation}
Let's compute $\psi Y^+_{1/2,1/2}$.
\be{eqnarray}
\{\psi Y^+_{1/2,1/2},y\}&=&
\psi Res_z\left(b_{-1}ce^{(iX+\phi)/\sqrt{2}}\right)(z)\cO_{1/2,-1/2}
\nonum\\
&=&\psi Res_z e^{(iX+\phi)/\sqrt{2}}(z)\cO_{1/2,-1/2}
\nonum\\
&=&1
\lb{3.22}\nnb\\
\{\psi Y^+_{1/2,1/2},x\}&=&
\psi Res_z e^{(iX+\phi)/\sqrt{2}}(z)\cO_{1/2,1/2}
\nonum\\
&=&0
\lb{3.23}
\end{eqnarray}
Therefore, we have
\be{equation}\lb{3.24}
\psi Y^+_{1/2,1/2}=\partial_y
\end{equation}
Similarly,
\be{equation}\lb{3.25}
\psi Y^+_{1/2,-1/2}=\partial_x
\end{equation}
This proves that $\psi$ maps onto \cA.

It is now clear that under $\psi$, the monomials
\be{equation}\lb{3.26}
{\left(\cO_{1/2,1/2}\right)}^n\cdot
{\left(\cO_{1/2,-1/2}\right)}^m\cdot
{\left(Y^+_{1/2,-1/2}\right)}^\nu\cdot
{\left(Y^+_{1/2,1/2}\right)}^\mu
\end{equation}
are sent to
\be{equation}\lb{3.27}
x^n\cdot y^m\cdot {\partial_x}^\nu\cdot {\partial_y}^\mu
\end{equation}
in \cA, where $n,m$ are nonnegative integers and $\nu,\mu$ are
0 or 1. Now by momentum counting and the multiplicity results
in Theorem \ref{thm3.1}, we see that the monomials \erf{3.26}
form a basis of $H(+)$. In particular, as a graded $\bullet$-algebra,
$H(+)$ is isomorphic to \cA.

This completes the proof of all of our claims.
To summarize, {\em we have an exact sequence of Gerstenhaber algebras:}
\be{equation}\lb{3.28}
0\lra H^*(-)\hra H^*\stackrel{\psi}{\lra} \cA^*\lra0
\end{equation}
{\em Moreover, there is a  splitting isomorphism
$\cA^*\stackrel{\sim}{\lra}H^*(+)$,
as associative algebras.}
To simplify notations,
we denote
\be{eqnarray}\lb{3.29}
x&=&\cO_{1/2,1/2}\nonum\\
y&=&\cO_{1/2,-1/2}\nonum\\
\partial_x&=&Y^+_{1/2,-1/2}\nonum\\
\partial_y&=&Y^+_{1/2,1/2}
\end{eqnarray}
It should become clear from the context when we write
$x,y,\px,\py$, whether they live in $H(+)$ or in \cA.

\subsection{The structure of the dot product}\lb{sec3.5}

The original goal of this chapter was to apply section \ref{sec2}
to understand the structure of the
Gerstenhaber algebra $H^*$. This means that we must at least
know how to describe the operations
$\bullet,\{,\}:H\times H\lra H$ in simple terms.
We will first focus on the dot product.

We have already fully understood
\be{equation}\lb{3.30}
\bullet:H(+)\times H(+)\lra H(+)
\end{equation}
Since we have established that $H(-)$ is an ideal in $H$,
the problem is further reduced to studying
\be{equation}\lb{3.31}
\bullet:H\times H(-)\lra H(-)
\end{equation}
i.e. studying $H(-)$ as a module over
the algebra $H$.
In the appendix, we show that
\be{equation}\lb{3.32}
H(-)\cdot H(-)=0
\end{equation}
To understand the dot product, it remains to study
$H(-)$ as a module over the $\bullet$-algebra $H(+)$.

Since $H(+)$ is a polynomial algebra, it is graded by
the degree. The subspace $H(+)[n]$ of polynomials of a fixed
degree $n$ is of course finite dimensional. Let
\be{equation}\lb{3.32a}
H(+)'=\oplus H(+)[n]'
\end{equation}
be the restricted dual of $H(+)$. Let
$\sigma$ be the linear anti-involution of $H(+)$ defined by
\be{eqnarray}\lb{3.33}
\sigma(x)&=&-x\nonum\\
\sigma(y)&=&-y\nonum\\
\sigma(\partial_x)&=&\partial_x\nonum\\
\sigma(\partial_y)&=&\partial_y\nonum\\
\sigma(u\cdot v)&=&\sigma(v)\cdot\sigma(u),\ \ for\ all\ u,v
\end{eqnarray}
Then we can define an $H(+)$-module structure on $H(+)'$ as follows:
for any linear functional $\lambda\in H(+)'$, and any $u,v\in H(+)$,
we let
\be{equation}\lb{3.34}
(u\cdot \lambda)(v)=\lambda(\sigma(u)\cdot v)
\end{equation}
We denote this dual module by $H(+)^\sigma$.
We claim that $H(-)$ is isomorphic to $H(+)^\sigma$
as $H(+)$-modules.

First recall that
on the the BRST complex,
 there is a non-degenerate bilinear pairing
\be{equation}\lb{3.35}
\lgl,\rgl : C^r(p,\alpha)\times C^{3-r}(-p,2\sqrt{2}-\alpha)\lra \bC
\end{equation}
which satisfies
\be{equation}\lb{3.35a}
\lgl\bone,\partial^2c\ \partial c\ ce^{2\sqrt{2}\phi}\rgl=2
\end{equation}
In terms of Figure 1, each point $(p,\alpha)$
of the lattice \cL is paired
with its image $(-p,2\sqrt{2}-\alpha)$
under the reflection through the point $(0,\sqrt{2})$.
With respect to this pairing, we have
\be{eqnarray}\lb{3.36}
\alpha_n^\dagger&=&-\alpha_{-n}\nonum\\
j_n^\dagger&=&-j_{-n}-2i\sqrt{2}\delta_{n,0}\nonum\\
c_n^\dagger&=&c_{-n}\nonum\\
b_n^\dagger&=&b_{-n}
\end{eqnarray}
This implies that the BRST operator $Q$ is self-adjoint, and hence
there is an induced nondegenerate pairing on the cohomology:
\be{equation}\lb{3.37}
\lgl,\rgl :H^r(p,\alpha)\times H^{3-r}(-p,2\sqrt{2}-\alpha)\lra \bC
\end{equation}

Notice that by definition \erf{3.15}, $H(+)$ pairs with $H(-)$
in a natural way. Thus we simply define the isomorphism
\be{equation}\lb{3.38}
H(-)\lra H(+)',\ \ \ \lambda\mapsto \lgl \lambda,\ \rgl
\end{equation}
The question is: does this map respect the action of $H(+)$?
i.e. do we get, for any $u,v$ in $H(+)$,
\be{equation}\lb{3.39}
\lgl u\cdot\lambda,v\rgl=\lgl \lambda, \sigma(u)\cdot v\rgl\ ?
\end{equation}
To answer, it is enough to check this for
the generators of $H(+)$: $u=x,y,\partial_x,\partial_y$.
We illustrate this in the simplest case $u=\partial_y$,
the rest being similar but more tedious.

By definition of the dot product \erf{2.7}, we have
\be{eqnarray}\lb{3.40}
\lgl \partial_y\cdot\lambda,v\rgl
&=& Res_z\lgl -ce^{(iX+\phi)/\sqrt{2}}(z)\lambda,v\rgl
\frac{1}{z}\nonum\\
&=& -\sum_n\lgl
c_{-n}{\left(e^{(iX+\phi)/\sqrt{2}}\right)}_n\lambda,v\rgl\nonum\\
&=& -\sum_n\lgl\lambda,
{\left(e^{(iX+\phi)/\sqrt{2}}\right)}_{-n}c_nv\rgl\nonum\\
&=& \lgl\lambda,\partial_y\cdot v\rgl
\end{eqnarray}
But because $\sigma(\partial_y)=\partial_y$, we see that
eqn \erf{3.39} indeed checks out for $u=\partial_y$.
The other cases are done in a similar way.
Thus we have
\be{equation}\lb{3.41}
H(-)\cong H(+)^\sigma
\end{equation}

\subsection{The bracket structure}\lb{sec3.6}

To understand the bracket structure, let's consider the  exact sequence
of graded Lie algebras:
\be{equation}\lb{3.41a}
0\lra H(-)\lra H\stackrel{\psi}{\lra} \cA\lra 0
\end{equation}
We would like to describe $H$ as an
extension of the (graded) Lie algebra \cA, by a (graded) module $H(-)$.
(The bracket on $H(-)$ is zero by Appendix A.)

The first ingredient for describing the extension is the two-cocycle
$\gamma$ corresponding to the sequence \erf{3.41a}.
This is a bilinear map $\gamma: \cA\times\cA\lra H(-)$
which can be  computed quite easily.
We will give the formula without going into the details. In terms of
the canonical basis $x^n\cdot y^m\cdot \px^\nu\cdot \py^\mu$ of \cA, we have
\be{equation}
\gamma(
x^n\cdot y^m\cdot \px^\nu\cdot \py^\mu,
x^{n'}\cdot y^{m'}\cdot \px^{\nu'}\cdot \py^{\mu'})
=\delta_{n,0}\delta_{m,0}\delta_{m',0}\delta_{n',0}
\{\px^\nu\cdot \py^\mu,
\px^{\nu'}\cdot \py^{\mu'}\}
\end{equation}
Note that the bracket on the right hand side is defined on $H$.
The right hand side is not identically zero because $\{\px,\py\}$ is
a non-zero element of $H(-)$ (see Appendix A).

We now describe the module $H(-)$. Since $\psi:H\lra\cA$ is a projection map,
every element in \cA has the form $\psi u$,
for some cohomology class $u$ in $H(+)$.
The action of $\cA$ on the module $H(-)$ is then defined by
\be{equation}
(\psi u)\lambda = \{u,\lambda\}
\end{equation}
where $\lambda$ is in $H(-)$. We claim that $H(-)$ is isomorphic
to the restricted dual module
$\cA^{\pi\circ\sigma}$
defined as follows.
Let $\cA^\pi$ be the twisted adjoint representation
given by
\be{equation}
\pi(a)b= (-1)^{|a|-1}\left( \{a,b\}-2(\Delta a)\cdot b\right)
\end{equation}
for $a,b$ in \cA.
Here $\Delta$ is defined by eqn \erf{3.52}.  Then
$\cA^{\pi\circ\sigma}$
is defined to be the $\sigma$-dual of $\cA^\pi$.
That is, $\cA'$ is the underlying space of
$\cA^{\pi\circ\sigma}$; and an element $a$ of \cA acts on
an element $\chi$ of $\cA'$ by
\be{equation}
\lgl a\chi,b\rgl=\lgl\chi,\pi\circ\sigma(a)b\rgl
\end{equation}

To prove the claim, define the linear isomorphism
\be{equation}
\phi: H(-)\lra\cA^{\pi\circ\sigma}=\cA'
\ \ \ \ \lambda\mapsto\lgl\lambda,\psi^{-1}(\cdot)\rgl
\end{equation}
Note that $\phi$ is independent of the choice of the ``inverse''.
We need to check that $\phi a\lambda= a\phi\lambda$, for all $a$ in \cA
and $\lambda$ in $H(-)$.
Equivalently, we can check that for all $u,v$ in $H(+)$,
\be{equation}
\lgl \phi(\psi u)\lambda,\psi v \rgl
=\lgl\phi\lambda,(\pi\circ\sigma(\psi u))\psi v\rgl
\end{equation}
It turns out that all we need to use is Lemma \ref{lem2.1},
the fact that under the pairing on $H$, $b_0^\dagger=b_0$, and that
the map $\psi$ intertwines $b_0$ with $-\Delta$ (see section \ref{sec3.8}):
\be{eqnarray}
\lgl \phi(\psi u)\lambda,\psi v \rgl
&=& \lgl  \{u,\lambda\}, v \rgl\nnb\\
&=&\lgl \lambda, (-1)^{|u|-1}\left(\{u^\dagger,v\}
+2(b_0u^\dagger)\cdot v\right)\rgl\nnb\\
&=&\lgl \lambda, (-1)^{|\psi u|-1}\psi^{-1}\left(\{\sigma(\psi u),\psi v\}
-2(\Delta\sigma(\psi u))\cdot \psi v\right)\rgl\nnb\\
&=&\lgl \lambda, \psi^{-1}\left(\pi\circ\sigma(\psi u)\right)(\psi v)\rgl\nnb\\
&=&\lgl\phi\lambda,(\pi\circ\sigma(\psi u))(\psi v)\rgl
\end{eqnarray}
This proves our claim.

Thus we have shown that {\em $H$, as a graded Lie algebra, is an
extension
of the Lie algebra of polyvector fields \cA by the dual module
$\cA^{\pi\circ\sigma}$. Moreover, this extension is characterized
by the two-cocycle $\gamma$ above.}

\subsection{The $b_0$ operator}\lb{sec3.7}

By virtue of Lemma \ref{lem2.1},
studying the operator
\be{equation}\lb{3.44}
b_0 : H^n\lra H^{n-1}
\end{equation}
should allow us to better understand the bracket product.
Let's first focus on $H(+)$.

Since there are no states of ghost number -1,
any polynomial $f$ in $H^0$ satisfies
\be{equation}\lb{3.45}
b_0f=0
\end{equation}
Similarly, we have for any two polynomials $f,g$,
\be{equation}\lb{3.46}
\{f,g\}=0
\end{equation}

By direct calculation using eqns \erf{3.29},  we also have
\be{eqnarray}\lb{3.47}
\{\partial_x,x\}&=&\{\partial_y,y\}=1\nonum\\
\{\partial_x,y\}&=&\{\partial_y,x\}=0\nonum\\
b_0\partial_x&=&b_0\partial_y=0
\end{eqnarray}
Thus by Lemma \ref{lem2.1}, we have
\be{eqnarray}\lb{3.48}
\partial_xf&=&\{\partial_x,f\}=-b_0(f\cdot\partial_x)\nonum\\
\partial_yf&=&\{\partial_y,f\}=-b_0(f\cdot\partial_y)
\end{eqnarray}
That is, $b_0$ acts on the 1-vector fields $H^1(+)$ by $-div$.

Finally, since $b_0$ kills both $\partial_x,\partial_y$,
we have (Lemma \ref{lem2.1})
\be{eqnarray}\lb{3.49}
\{\partial_x,\partial_y\}&=&-b_0(\partial_x\cdot\partial_y)\nonum\\
&=&-b_0(-\partial c\ ce^{\sqrt{2}\phi})\nonum\\
&=& ce^{\sqrt{2}\phi}
\end{eqnarray}
On the 2-vector fields $H^2(+)$, we have
\be{eqnarray}\lb{3.50}
b_0(f\cdot\partial_x\cdot\partial_y)&=&
\{f,\partial_x\cdot\partial_y\}
+f\cdot b_0(\partial_x\cdot\partial_y)\nonum\\
&=&\partial_yf\cdot\partial_x-\partial_xf\cdot\partial_y
-f\cdot \{\partial_x,\partial_y\}
\end{eqnarray}
This completes the description of the $b_0$ operator
on $H(+)$.

Note that because $\{\partial_x,\partial_y\}\in H(-)$, neither
the $b_0$ operator nor the bracket stabilizes $H(+)$.
However, since $b_0$ carries zero momenta, the definition
\erf{3.15} of $H(-)$ shows that $b_0$ preserves $H(-)$.
Now because the dot product restricted to $H(-)$ is zero
(see Appendix A), it follows that the bracket restricted to
$H(-)$ is also zero.

Since $b_0^\dagger=b_0$ under the pairing
between $H(+)$ and $H(-)$, we have
$\lgl b_0\lambda,v\rgl =\lgl\lambda,b_0v\rgl$ for all
$\lambda$ in $H(-)$ and $v$ in $H(+)$.

%
%
\subsection{Relation between $b_0$ and $\Delta$}\lb{sec3.8}

As remarked in section \ref{sec2}, $b_0$ plays a role
analogous to that
of certain differential operator $\Delta$, in the anti-bracket
formalism (see \cite{W2}). If we now consider the latter
in the case where the field space is the xy-plane, then
$\Delta$ is acting in a certain
algebra of polyvector fields on the plane. Here we take
the algebra to be $\cA=\bC[x,y,\px,\py]$. Then we have
\be{equation}\lb{3.52}
\Delta=\frac{\partial \ }{\partial x}
\frac{\partial \ }{\partial x^*}
+\frac{\partial \ }{\partial y}
\frac{\partial \ }{\partial y^*}
\end{equation}
where $x^*=\px, y^*=\py$.
We claim  that the homomorphism $\psi$
of Gerstenhaber algebras intertwines between $b_0$ and $-\Delta$,
i.e. for every BRST class $u$, we have
\be{equation}\lb{3.53}
\psi(b_0u)=-\Delta(\psi u)
\end{equation}

Since $b_0$ stabilizes $H(-)$ which is the kernel of $\psi$,
eqn \erf{3.53} holds trivially when $u$ is in $H(-)$. So
let's focus on $H(+)$.
As before, $b_0$ kills $H^0$. On the other hand, $\Delta$ kills
$\cA^0$. On $H^1(+)$, $-b_0$ acts as the divergence operator \erf{3.48}.
But so does $\Delta$ on $\cA^1$.
Finally, $b_0$ acts on $H^2(+)$ by eqn \erf{3.50}.
Since $\{\px,\py\}$ of $H$ gets sent to zero by $\psi$, we have
\be{equation}\lb{3.54}
\psi b_0(f\cdot\px\cdot\py)=\py f\cdot x^*-\px f\cdot y^*
\end{equation}
But the right hand side coincides with $-\Delta(f\cdot x^*\cdot y^*)$.
This completes the proof of eqn \erf{3.53}.

To summarize our application of section 2 to the c=1 model, we have

\be{thm}\lb{thm3.2} Let $H^*$ be the chiral cohomology of the c=1 model.
Let $\cA^*$ be the Gerstenhaber algebra
$\bC[x,y,\partial_x,\partial_y]$ (see Appendix B).
Then the following holds:\\
(a) There is an exact sequence of
Gerstenhaber algebras
\[
0\lra H(-)\lra H\lra \cA\lra 0
\]
where $H(-)$ is an ideal in which both products are identically zero.\\
(b) $H^*(+)$ is closed under the dot product, and it is canonically isomorphic
to $\cA^*$, as an associative algebra.\\
(c) Let $H(+)^\sigma$ be the restricted dual of $H(+)$
defined by  the $\bullet$-algebra anti-involution $\sigma$
(section \ref{sec3.5}).
Then the ideal $H(-)$ is isomorphic, as module over $H(+)$, to $H(+)^\sigma$.\\
(d) $H^*(+)$ is {\em not} closed under the bracket product. The
sequence in (a) does not split as an exact sequence of graded Lie algebras
(For more details on the bracket, see section \ref{sec3.6}).\\
(e) The projection map $H\lra\cA$ intertwines $-b_0$ on $H$
and the differential operator
\[
\Delta=\frac{\partial \ }{\partial x}
\frac{\partial \ }{\partial x^*}
+\frac{\partial \ }{\partial y}
\frac{\partial \ }{\partial y^*}
\]
on \cA where $x^*=\px, y^*=\py$.
\end{thm}

\subsection{Discussion}

\subsubsection{
Additional examples of string backgrounds}

So far we have discussed in detail the algebraic structure of the
c=1 model. We will now make some remarks about the 26 dimensional
bosonic string background.

In reference \cite{FGZ}, the authors consider the string background
consisting of 26 bosons compactified on a torus. This may be viewed
as a chiral algebra constructed from a 26 dimensional Lorentzian lattice.
The chiral BRST cohomology $H$
in this case has the following structure: the
ground ring $H^0$ consists of only the identity operator.
It is also known that the space $P$ of old physical states
is isomorphic to $H^1$. We show, in section \ref{sec2.4}, that
this isomorphism is in fact at the level of Lie algebras, i.e.
the Lie bracket on $P$ is compatible with the Gerstenhaber bracket on $H^1$.
The ghost number two cohomology $H^2$ is a module over $H^1$. Since
$\{H^2,H^2\}\subset H^3$ and $H^3$ is one-dimensional, the bracket provides
a bilinear form on $H^2$ which is invariant under the $H^1$-action.
We believe that this new structure is worthy of further study.

In reference \cite{LZ3}, we study the backgrounds in which
the $c<1$ minimal models are coupled to the Liouville field
from the BRST point of view. Since then, many physicists  have studied
the dot product structure of the BRST cohomology. But we will not attempt
to review the recent developments (see for example \cite{KS}\cite{KMS}
for references).
The full structure of the Gerstenhaber
algebras for $c<1$ has not been worked out. Note that we will
need some modifications to our theory to take into account the
operators with non-integral dimension in the $c<1$ theories.

\subsubsection{Deformations of a chiral algebra?}\lb{sec3.9.2}

In a recent discussion with Greg Moore, we have learned that
deformations of a conformal field theory ought to be connected to
BRST invariant operators. The so-called marginal operators correspond
to first order deformations (perturbations) of a fixed CFT.
The so-called exactly marginal operators give rise to deformations
to all orders of the CFT. These deformations occur in the context
of a full two-sided CFT. We propose that one should also consider
deformations of the corresponding {\em chiral} theory.

Given a fixed chiral background $V$ and a dimension zero
 BRST invariant operator $\phi(z)$, we should consider the
associated current $(b_{-1}\phi)(z)$.
The first measure for the failure of
$(b_{-1}\phi)(z)$ to be ``marginal'' should be
indicated by the OPE with itself. Specifically, the first order pole
in $(b_{-1}\phi)(z)(b_{-1}\phi)(w)$
should represent an obstruction for $(b_{-1}\phi)(z)$ to be marginal.
\be{pro} The first order pole of the OPE
$(b_{-1}\phi)(z)(b_{-1}\phi)(w)$
vanishes if
and only if
the Gerstenhaber bracket $\{\phi,\phi\}$ vanishes.
\end{pro}
The above necessary and sufficient
condition is very reminiscent of the classical
BV equation (on the physical side) \cite{W2}, and the condition for second
order
deformation of an associative algebra (on the mathematical side) \cite{Gers2}.
It also reminds us of the existence condition for a Poisson structure
on a manifold (see Proposition \ref{pro1.2}).

In the c=1 model, let's try to solve the equation
\be{equation}\lb{3.55}
\{\phi,\phi\}=0
\end{equation}
We focus on $\phi\in H^*(+)$. First observe that by graded skew
symmetry of the bracket, the  solution set of eqn \erf{3.55} is
invariant under the translation by any BRST class $\psi$ with
odd ghost number. That is, if $\phi$ is a solution then so is
$\phi+\psi$. Since nontrivial states in the c=1 model only have ghost
number zero through three, it is enough to consider solutions
involving ghost number zero and two:
\be{equation}
\phi=f+g\cdot\px\cdot\py
\end{equation}
where $f,g$ are polynomial functions to be determined.
It is easy to see that eqn \erf{3.55} is now equivalent to
the following two equations:
\be{eqnarray}
\{f,g\cdot\px\cdot\py\}&=&0\nnb\\
\{g\cdot\px\cdot\py
,g\cdot\px\cdot\py\}&=&0
\end{eqnarray}
Both equations are easy to solved. The most general
solution to eqn \erf{3.55} in $H^*(+)$
is of one of the following two types (up to translation
by odd ghost number state):\\
(i) $\phi=f$ for any polynomial function $f$;\\
(ii) $\phi=const. + g\cdot\px\cdot\py$ for any
polynomial function $g$ with zero constant term.\\

\subsubsection{
Modules over the BRST algebra}

As in the case of Hochschild cohomology \cite{Gers2}, our theory
of Gerstenhaber algebra can be generalized to the case of modules.
More precisely, given a chiral background $V$ and a $V$-module $M$
(for definition, see \cite{FHL}), we can form the corresponding
BRST complex:
\be{equation}
M\otimes \Lambda^*
\end{equation}
It can be shown that the correponding cohomology $H^*(M)$ is a
module over the Gerstenhaber algebra $H^*(V)$. This is analogous
to the situation in Hochschild theory.

When we pass to BRST cohomology,
the space of intertwiners of $V$-modules descends to a space
of intertwiners of $H^*(V)$-modules.

\subsubsection{
Topological chiral algebras}

We can significantly generalize the notion of a string background to
the notion of a topological conformal field theory (see for example
\cite{DVV}\cite{EguYa} and references therein).
\be{dfn}
A topological chiral algebra (TCA) consists of the following data:
a super chiral algebra $C^*$,
a weight one even
current $F(z)$ whose charge $F_0$ is the fermion number operator,
a weight one odd primary field $J(z)$ having fermion number one and
having a square zero charge $Q$,
and a weight two odd primary field $G(z)$
having fermion number -1  and satisfying $[Q,G(z)]=L(z)$
where $L(z)$ is the stress-energy field.
We denote the cohomology of the complex $(C^*,Q)$ by $H^*(C)$.
\end{dfn}
Remarkably, all of the structures of our theory
generalize to the case of topological chiral algebras.
In particular, if we replace the BRST complex by a general
TCA $C^*$,
the ghost number current $c(z)b(z)$ by $F(z)$,
the BRST current by $J(z)$, the BRST operator by the
charge $Q$ of $J(z)$, and the anti-ghost $b(z)$ by $G(z)$,
then the exact translation of Theorem \ref{thm2.2} holds for $H^*(C)$.
Moreover, the appropriate translations of all the statements
in sections \ref{sec2.1}---\ref{sec2.3} hold true in this general context.
In particular, we have a coboundary Gerstenhaber algebra on $H^*(C)$
(see Definition \ref{def5.3}), and up to homotopy on $C^*$.

Our generalization incorporates many interesting examples.
The so-called N=2 twisted super conformal field theories
are known to give rise to examples of TCA's
(see for example \cite{DVV}\cite{EguYa}).
It can be shown that in these examples, the $G_0$ operator
acts by zero on cohomology provided the N=2 theory is unitary.
As a consequence, by (translation of) Lemma \ref{lem2.1},
the Gerstenhaber bracket is identically zero in cohomology.

Even in the general context of TCA's, the question we raise in
section \ref{sec3.9.2} still makes sense. In particular,
what is the relation between the equation $\{\phi,\phi\}=0$
and deformations of $C^*$? More generally, if $\hbar$ is a
parameter, what is the meaning (physical and mathematical) of
the ``quantum BV'' equation:
\be{equation}
\hbar G_0\phi+\{\phi,\phi\}=0 \ \ ?
\end{equation}

\subsubsection{ Closed string (field) theory}

Up to now, we have restricted our discussion to {\em chiral} field
theories. In the case of string theory, chiral theories alone
are not adequate for describing closed strings \cite{WZ}\cite{Zw}.
At the algebraic level, there are
at least two additional things we must do.

First we must tensor the left and the right moving BRST complexes:
\be{equation}
C^*\otimes \bar{C}^*
\end{equation}
We observe that this double complex admits all the
interesting algebraic structures that the chiral sectors have.
However, while the Gerstenhaber bracket on each of the chiral
sectors satisfies the Jacobi identity off-shell, the bracket on
the double complex does so {\em only} up to homotopy.
Anyway, the cohomology of the double complex is again
a coboundary Gerstenhaber algebra (CGA)(Definition \ref{def5.3})
 which is the tensor product
of the left and the right CGA's. The ``coboundary operator''
is given by $\Delta=b_0+\bar{b}_0$.
Off-shell, the double complex is a CGA up to homotopy.

However, it seems that the proper counterpart of the $b_0$
operator is not $b_0+\bar{b}_0$ in closed string theory \cite{WZ},
but rather it should be $b_0^-=b_0-\bar{b}_0$. This means that
we should also twist the Gerstenhaber algebra structure on
the double complex above by replacing
$b_0+\bar{b}_0$ by
$b_0-\bar{b}_0$. This is equivalent to twisting the Gerstenhaber
bracket on $\bar{C}^*$ by a minus sign.
Thus on the double complex, the twisted bracket is given by
\be{equation}\lb{3.56}
(-1)^{|u|}\{u,v\}=b^-_0(u\cdot v)-(b_0^-u)\cdot v
-(-1)^{|u|}u\cdot(b_0^- v)
\end{equation}
where $u,v$ are elements of the double complex.
Note that the dot product on the double complex remains
compatible with this new bracket.

The notion of a strongly homotopy Lie algebra is well known
in mathematics (see Lada-Stasheff's recent review
\cite{LS} for an introduction).
Recently, this structure  has arisen in the context of
closed string theory \cite{WZ}.
A similar structure also appears in closed string field theory
--- both on-shell and off-shell \cite{Zw}. How are the off-shell
SHLA in \cite{WZ}\cite{Zw} related to the HLA defined by eqn \erf{3.56}?
Are there any {\em higher} homotopies (see \cite{LS} for definition)
in connection with (a)-(e) of our Theorem \ref{thm2.2}?
Our work searching for such higher homotopies is underway.
We remark that Witten-Zwiebach's construction
of the Lie bracket involves only three-point functions,
hence may be treated algebraically --- just as our construction
in eqn \erf{3.56}.
However, their construction of the higher homotopies requires
the consideration of the geometry of moduli spaces. It is
therefore interesting to find a precise connection between
the the geometric approach and our algebraic approach.

%

\setcounter{equation}{0}

\section{Appendix A}

The main purpose of this section is to show that $H^*$
is generated, as a Gerstenhaber algebra, by four generators
$x,y,\partial_x,\partial_y$. Moreover, we describe
a basis of $H(-)$, which is dual and complementary to the
basis of $H(+)$ given in \erf{3.26},
and hence show that $H(-)$ is an ideal with one
generator $\{\px,\py\}$. The pairing between
the two bases is $sl_2$ invariant.
We show, as a consequence, that $H(-)$ is a subalgebra
of $H$ with the zero product. This last assertion can
also be drawn from \cite{WuZh}, where the product
is explicitly computed.
We also describe the action of $b_0,x,y,\px,\py$
on $H(-)$ in terms of the above dual basis.

\subsection{The dual basis}

Using eqns \erf{3.21},\erf{3.29}, it is easy to get
\be{eqnarray}\lb{a.1}
\{\px,\py\}&=&ce^{\sqrt{2}\phi}\nonum\\
\px\cdot\py&=&-\partial c\ ce^{\sqrt{2}\phi}
\end{eqnarray}
With respect to the pairing \erf{3.37}, we have
\be{equation}\lb{a.2}
\lgl\px\cdot\py,\{\px,\py\}\rgl=-1
\end{equation}
Since $\px,\py$ are self-adjoint, it follows
that the four states
$\{\px,\py\}$, $\px\cdot\{\px,\py\}$, $\py\cdot\{\px,\py\}$
and $\px\cdot\py\cdot\{\px,\py\}$ are paired with $-\px\cdot\py$,
$-\py$, $\px$, and \bone respectively.

To simplify notations, we sometimes write $\{\px^n u$ to mean
applying the bracket of $\px$ with $u$,
n times: $\{\px,...,\{\px,u\}...\}$. Similarly for $\{\py^n u$.

Since $H(-)$ is an ideal containing
$\{\px,\py\}$, it follows that $x\cdot\{\px,\py\}$ must lie in $H(-)$.
The state $x\cdot\{\px,\py\}$ has quantum numbers $(p,\alpha,gh\#)
=(\sqrt{2}/2,\sqrt{2}/2,1)$. According
to eqn \erf{3.15}, this state must be zero.
Thus we get, for any polynomial $f$ without a constant term,
\be{equation}\lb{a.2a}
f\cdot\{\px,\py\}=0
\end{equation}
Expanding $0=\{\px, x\cdot\{\px,\py\}\}$, we get
\be{equation}\lb{a.3}
x\cdot\{\px,\{\px,\py\}\}=-\{\px,x\}\cdot\{\px,\py\}=-\{\px,\py\}
\end{equation}
Applying the same trick repeatedly, we get
\be{equation}\lb{a.4}
x\cdot\{\px^n \{\py^m \{\px,\py\}
=-n\{\px^{n-1} \{\py^m \{\px,\py\}
\end{equation}
Similarly, we have
\be{equation}\lb{a.5}
y\cdot\{\px^n \{\py^m \{\px,\py\}
=-m\{\px^n \{\py^{m-1} \{\px,\py\}
\end{equation}
This shows that the $\{\px^n\{\py^m\{\px,\py\}$ are all nonzero
because we can hit them with $x,y$ repeatedly to get down
to a nonzero multiple of $\{\px,\py\}$.
Moreover, they all have distinct quantum numbers, and hence
are linearly independent. In fact, by comparing them with the
quantum number spectrum given by Theorem \ref{thm3.1},
we see that they form a basis of $H^1(-)$.

We can repeat the above argument, with $\{\px,\py\}$ replaced by
$\px\cdot\{\px,\py\}$, $\py\cdot\{\px,\py\}$ or
$\px\cdot\py\cdot\{\px,\py\}$. Then eqns \erf{a.4},\erf{a.5}, with
the appropriate changes, still hold. Again we get some bases
for $H^2(-)$ and $H^3(-)$. In fact, by comparing quantum numbers,
one can easily see that the new bases we obtained
can be identified with the bases introduced in \cite{WZ}.
The identification goes as follows:
for nonnegative half integer $s$, and $n=-s,-s+1,...,s$,
\be{eqnarray}
Y^-_{s,n}&\sim&\{\px^{s-n}\{\py^{s+n}\left(\{\px,\py\}\right)
\lb{a.6}\\
aY^-_{s,n}&\sim&(s-n)\{\px^{s-n-1}\{\py^{s+n}
\left(\px\cdot\{\px,\py\}\right)\nnb\\
& &+(s+n)\{\px^{s-n}\{\py^{s+n-1}
\left(\py\cdot\{\px,\py\}\right)
\lb{a.7}\\
P_{s,n}&\sim&\{\px^{s-n+1}\{\py^{s+n}
\left(\py\cdot\{\px,\py\}\right)\nnb\\
& &-\{\px^{s-n}\{\py^{s+n+1}
\left(\px\cdot\{\px,\py\}\right)
\lb{a.8}\\
aP_{s,n}&\sim&\{\px^{s-n}\{\py^{s+n}
\left(\px\cdot\py\cdot\{\px,\py\}\right)
\lb{a.9}
\end{eqnarray}
This shows that $H(-)$, as an ideal of the Gerstenhaber algebra
$H$, is generated by $\{\px,\py\}$.
For completeness, let's write down a basis for $H(+)$ as well:
\be{eqnarray}
aY^+_{s,n}&\sim&x^{s-n}\cdot y^{s+n}\cdot\px\cdot\py
\lb{a.10}\\
Y^+_{s,n}&\sim&\px(x^{s-n}\cdot y^{s+n})\cdot\py
-\py(x^{s-n}\cdot y^{s+n})\cdot\px
\lb{a.11}\\
a\cO_{s,n}&\sim&x^{s-n}\cdot y^{s+n}\cdot(x\cdot\px+y\cdot\py)
\lb{a.12}\\
\cO_{s,n}&\sim&x^{s-n}\cdot y^{s+n}
\lb{a.13}
\end{eqnarray}

\subsection{The $sl_2$ action}

What can we say about these bases?
Already it has been indicated in \cite{W}\cite{WZ} that
for fixed $s$, each of the multiplets ($n=-s,-s+1,...,s$) in
eqns \erf{a.10}-\erf{a.13} is an $sl_2$ spin $s$ multiplet.
One can represent the $sl_2$ generators
simply by $\{x\cdot\py,*\}$, $\{x\cdot\px-y\cdot\py,*\}$
and $\{y\cdot\px,*\}$. Computing the action of these $sl_2$
generators on \erf{a.10}-\erf{a.13} is straightforward,
because the relations \erf{3.47} \erf{a.2a}
suffice.
But computing the action on those multiplets in
eqns \erf{a.6}-\erf{a.9} is more difficult
because the relations \erf{3.47} \erf{a.2a} alone are not enough.

Fortunately, we can once again use the pairing between
$H(+)$ and $H(-)$.
It is easy to check that for any $u,v$ in $H$, and any
one of the $sl_2$ generators $\{X,*\}$ above, we have
\be{equation}\lb{a.14}
\lgl \{X,u\},v\rgl=\lgl u,-\{X,v\}\rgl
\end{equation}
i.e. the pairing is $sl_2$ invariant.
To see that each of the multiplets in the $H(-)$ sector
--- eqns \erf{a.6}-\erf{a.9}
(fixed $s$ and $n=-s,-s+1,...,s$) --- is indeed an $sl_2$ spin $s$
multiplet, it is enough to show that it pairs with a
multiplet of the same spin in the $H(+)$ sector.
Now using the adjointness property \erf{3.33} of
$x,y,\px,\py$, and using the dot products of $x,y$ with
each of the states in eqns \erf{a.6}-\erf{a.9}, we obtain
the orthogonality relations:
\be{eqnarray}
& &\lgl x^{s-n}\cdot y^{s+n}\cdot\px\cdot\py \ {\bf,} \
\{\px^{s-n}\{\py^{s+n}\left(\{\px,\py\}\right)\rgl\nnb\\
& &=-(s-n)!(s+n)!
\lb{a.15}\\
& &\lgl \px(x^{s-n}\cdot y^{s+n})\cdot\py
-\py(x^{s-n}\cdot y^{s+n})\cdot\px \ {\bf,}\nnb\\
& &(s-n)\{\px^{s-n-1}\{\py^{s+n}
\left(\px\cdot\{\px,\py\}\right)
+(s+n)\{\px^{s-n}\{\py^{s+n-1}
\left(\py\cdot\{\px,\py\}\right)\rgl\nnb\\
& &=-2s(s-n)!(s+n)!
\lb{a.16}\\
& &\lgl x^{s-n}\cdot y^{s+n}\cdot(x\cdot\px+y\cdot\py) \ {\bf,}\nnb\\
& &\{\px^{s-n+1}\{\py^{s+n}
\left(\py\cdot\{\px,\py\}\right)
-\{\px^{s-n}\{\py^{s+n+1}
\left(\px\cdot\{\px,\py\}\right)\rgl\nnb\\
& &=(2s+2)(s-n)!(s+n)!
\lb{a.17}\\
& &\lgl x^{s-n}\cdot y^{s+n} \ {\bf,} \
\{\px^{s-n}\{\py^{s+n}
\left(\px\cdot\py\cdot\{\px,\py\}\right)\rgl\nnb\\
& &=(s-n)!(s+n)!
\lb{a.18}
\end{eqnarray}
All the other inner products are zero. This proves our assertion.

\subsection{The dot product in $H(-)$}

Since there is no state with ghost number
greater than 3, the only possibly nonzero dot products
that we need to consider are those
on $H^1(-)\times H^1(-)$ and $H^1(-)\times H^2(-)$.

Consider the product in $H^1(-)$, which is spanned by the $Y^-_{s,n}$.
Obviously, the $sl_2$ action considered above acts by derivations
of the dot product. This means that the product of two states
$Y^-_{s,n}, Y^-_{s',n'}$ (see eqn \erf{a.6}),
must lie in the tensor product representation
of those two spins. In particular,
it must be a linear combination of states with spin $s''$ satisfying
$|s-s'|\leq s''\leq s+s'$. Now with this restriction and
by the conservation of quantum numbers
(i.e. $p,\alpha,gh\#$),  we must have
\be{equation}\lb{a.19}
Y^-_{s,n}\cdot Y^-_{s',n'}=constant \ P_{s+s',n+n'}
\end{equation}
To prove that this is zero, it is enough to show that
the inner product of the left hand side with
$a\cO_{s+s',n+n'}$ (see eqn \erf{a.12}) is zero.
Consider
\be{eqnarray}\lb{a.20}
& &\lgl a\cO_{s+s',n+n'},
Y^-_{s,n}\cdot Y^-_{s',n'}\rgl\nnb\\
& &\sim\lgl
x^{s+s'-n-n'+1}\cdot y^{s+s'+n+n'}\cdot\px,
Y^-_{s,n}\cdot Y^-_{s',n'}\rgl\nnb\\
& &+\lgl
x^{s+s'-n-n'}\cdot y^{s+s'+n+n'+1}\cdot\py,
Y^-_{s,n}\cdot Y^-_{s',n'}\rgl
\end{eqnarray}
Since $x,y$ are  anti-selfadjoint, we can bring their monomials
to the second slot, with only some sign change.
Because of eqn \erf{a.4} and the fact that $s+s'-n-n'+1> s-n$,
we have $x^{s+s'-n-n'+1}\cdot Y^-_{s,n}=0$.
Similarly, $y^{s+s'+n+n'+1}\cdot Y^-_{s',n'}=0$.
This shows that $H^1(-)\cdot H^1(-)=0$. The argument
for $H^1(-)\cdot H^2(-)=0$ is the same.

%
%
\subsection{$b_0,x,y,\px,\py$ acting on $H(-)$}

Finally, we would like to give a more explicit description
of $H(-)$ as a module over the Gerstenhaber algebra $H$.
We should therefore describe -- in terms of a nice basis -- how the generators
$x,y,\px,\py$ of $H$ act on $H(-)$, first by the dot product,
then by the bracket.

The actions of $x,y$ on the basis \erf{a.6}-\erf{a.9} of $H(-)$
have already been alluded to above. We summarize them as follows.
Let
$\lambda$ be one of the following four special states
$\{\px,\py\},\px\cdot\{\px,\py\},\py\cdot\{\px,\py\}$
or $\px\cdot\py\cdot\{\px,\py\}$. Then we have
\be{eqnarray}\lb{a.21}
x\cdot\{\px^n\{\py^m \lambda &=&
-n\{\px^{n-1}\{\py^m \lambda\nnb\\
y\cdot\{\px^n\{\py^m \lambda &=&
-m\{\px^n\{\py^{m-1} \lambda
\end{eqnarray}

To work out the action of $\px$, we note for example that
\be{eqnarray}\lb{a.22}
\px\cdot\{\py,\lambda\}&=&
\{\py,\px\cdot\lambda\}-\{\py,\px\}\cdot\lambda\nnb\\
&=&
\{\py,\px\cdot\lambda\}
\end{eqnarray}
This holds because $\{\py,\px\}$ and $\lambda$
are both in $H(-)$, and hence have zero dot product.
Similarly, we have
\be{eqnarray}\lb{a.23}
\px\cdot\{\px^n\{\py^m \lambda &=&
\{\px^n\{\py^m (\px\cdot\lambda)\nnb\\
\py\cdot\{\px^n\{\py^m \lambda &=&
\{\px^n\{\py^m (\py\cdot\lambda)
\end{eqnarray}

In section \ref{sec2.3}, we proved that $b_0$ acts as a derivation
of the bracket in $H$. Using this and the fact that
 $b_0$ kills both $\px,\py$, we get
\be{eqnarray}\lb{a.24}
b_0\{\px^n\{\py^m (\{\px,\py\})&=&0\nnb\\
b_0\{\px^n\{\py^m (\px\cdot\{\px,\py\})&=&
-\{\px^{n+1}\{\py^m (\{\px,\py\})\nnb\\
b_0\{\px^n\{\py^m (\py\cdot\{\px,\py\})&=&
-\{\px^n\{\py^{m+1} (\{\px,\py\})\nnb\\
b_0\{\px^n\{\py^m (\px\cdot\py
\cdot\{\px,\py\})&=&
-\{\px^{n+1}\{\py^m (\py\cdot\{\px,\py\})\nnb\\
& &+\{\px^n\{\py^{m+1} (\px\cdot\{\px,\py\})
\end{eqnarray}
We can now summarize the bracket operations of $x,y,\px,\py$
with $H(-)$:
\be{eqnarray}\lb{a.25}
\{\px,\left(\{\px^n\{\py^m \lambda\right)\}&=&
\{\px^{n+1}\{\py^m \lambda\nnb\\
\{\py,\left(\{\px^n\{\py^m \lambda\right)\}&=&
-\{\px^n\{\py^{m+1} \lambda\nnb\\
\{x,\left(\{\px^n\{\py^m (\{\px,\py\})\right)\}&=&0\nnb\\
\{y,\left(\{\px^n\{\py^m (\{\px,\py\})\right)\}&=&0\nnb\\
\{x,\left(\{\px^n\{\py^m (\px\cdot\{\px,\py\})\right)\}&=&
-\{\px^n\{\py^m (\{\px,\py\})\nnb\\
\{y,\left(\{\px^n\{\py^m (\px\cdot\{\px,\py\})\right)\}&=&0\nnb\\
\{x,\left(\{\px^n\{\py^m (\py\cdot\{\px,\py\})\right)\}&=&0\nnb\\
\{y,\left(\{\px^n\{\py^m (\py\cdot\{\px,\py\})\right)\}&=&
-\{\px^n\{\py^m (\{\px,\py\})\nnb\\
\{x,\left(\{\px^n\{\py^m (\px\cdot\py\cdot\{\px,\py\})\right)\}&=&
-\{\px^n\{\py^m (\py\cdot\{\px,\py\})\nnb\\
\{y,\left(\{\px^n\{\py^m (\px\cdot\py\cdot\{\px,\py\})\right)\}&=&
\{\px^n\{\py^m (\px\cdot\{\px,\py\})\nnb\\
\end{eqnarray}
where $\lambda$ is any one of the four special states.

\setcounter{equation}{0}

\section{Appendix B}

In this section, we will make a few introductory comments about the
mathematical theory of Gerstenhaber algebras; for a more mathematically
advanced and complete account of the theory see the forthcoming preprint
\cite{Zuck}.
\be{dfn}
A Gerstenhaber algebra $G^*$ is a \bZ-graded vector space equipped
with two bilinear multiplication operations, denoted by $u\cdot v$
and $\{u,v\}$ respectively, and satisfying the following assumptions:\\
(i) If $u$ and $v$ are homogeneous elements of degree $|u|$ and $|v|$
respectively, then $u\cdot v$ is homogeneous of degree $|u|+|v|$ and
$\{u,v\}$ is homogeneous of degree $|u|+|v|-1$.\\
(ii) Identities (a) through (e) from Theorem \ref{thm2.2} above
hold for any triple of homogeneous elements $u,v$ and $t$ in $G^*$.
\end{dfn}

{\bf Note:} We will call the product $u\cdot v$ the dot product
and the product $\{u,v\}$ the Gerstenhaber bracket product.

{\bf Remarks:}\\
(i) Physicists will want to call the degree of a homogeneous element the
ghost number. Mathematicians are primarily familiar with examples in which the
degree is bounded from above or from below by zero.\\
(ii) With respect to the dot product, $G^*$ is a \bZ-graded
supercommutative associative algebra.\\
(iii) Let $\Pi G^*$ be the graded vector space defined by
$\Pi G^k=G^{k-1}$. Then, with respect to the bracket product,
$\Pi G^*$ is a \bZ-graded Lie superalgebra.\\
(iv) Identity (e) in Theorem \ref{thm2.2} establishes the crucial
link between the two different products in $G^*$.\\

The following are a few classes of examples of a Gerstenhaber algebra.

A). The simplest Gerstenhaber algebras are the following: let
$A^*_n$ be the \bZ-graded supercommutative algebra generated by
$n$ variables $x_1,x_2,...,x_n$, of degree zero, and $n$ variables
$\partial_{x_1},\partial_{x_2},...,\partial_{x_n}$, of degree one.
We refer to an element of this algebra as a polyvector field.
The elements of degree zero are functions, the elements of degree
one are vector fields, of degree two are bivector fields, and so on.
The dot product is simply the graded commutative multiplication of
polyvector fields.

Long ago, Schouten \cite{Sch} defined a bracket operation on polyvector
fields (he thought of such fields as antisymmetric contravariant tensor
fields.) The Schouten bracket is characterized by the following:\\
(i) For any two function $f$ and $g$, $\{f,g\}=0$.\\
(ii) If $f$ is a function and $X$ is a vector field,
$\{X,f\}=-\{f,X\}=Xf$, i.e. the evaluation of the vector field
$X$ on the function $f$.\\
(iii) If $X$ and $Y$ are vector fields, then $\{X,Y\} =[X,Y]$,
the Lie bracket of the vector fields.\\
(iv) Together, the dot product and the Schouten bracket endow
$A^*_n$ with structure of a Gerstenhaber algebra. In particular,
identity (e) holds.\\

{\bf Remarks:} \\
(i) It is an elementary and useful exercise to write an explicit formula
for the Schouten bracket of any two polyvector fields.\\
(ii) The algebra $A_n^*$ appears in Witten's article \cite{W2},
where the variables $\partial_{x_k}$ are denoted by $x_k^*$, $k=1,...,n$,
and the Schouten bracket is defined as a kind of Poisson bracket with unusual
signs --- more precisely, an odd-Poisson bracket. (The
sign convention we use here differs slightly from the one
in \cite{W2}.) Thus the algebra $A_n^*$ arises naturally in the
theory of Batalin and Vilkovisky, who
employ a field space generalization of the algebras $A^*_n$.
We will say more about $A^*_n$ and BV theory below.

B). For a different class of Gerstenhaber algebras, let $g$ be any
Lie algebra, and let $\Lambda^*g$ be the Grassmann algebra
generated by $g$. We can define a bracket $\{,\}$ on $\Lambda^*g$
by requiring the following:\\
(i) If $a$ and $b$ are scalars, $\{a,b\}=0$.\\
(ii) If $X$ is in $g$ and $a$ is a scalar, then $\{X,a\}=0$.\\
(iii) If $X$ and $Y$ are in $g$, then $\{X,Y\}=[X,Y]$, the Lie
bracket in $g$.\\
(iv) The wedge product together with the bracket product endow
$\Lambda^*g$ with the structure of a Gerstenhaber algebra. In particular,
identity (e) above holds.\\

{\bf Remarks:}\\
(i) The algebra $\Lambda^*g$ is discussed briefly in Drinfeld's famous
article on quantum groups \cite{Drin}. However, Drinfeld mistakenly
implies that if $V_n$ is the Lie algebra of polynomial coefficient vector
fields in $n$ variables, then $\Lambda^*V_n$ is isomorphic as
a Gerstenhaber algebra to $A_n^*$. In fact, there is a Gerstenhaber
algebra homomorphism from $\Lambda^*V_n$ to $A^*_n$ which is an
isomorhism at the degree one level but which clearly fails to
be an isomorphism at degree zero, since $\Lambda^0V_n$ is just
one dimensional.\\
(ii) It is an elementary exercise to write the explicit bracket
of two exterior forms in the algebra $\Lambda^*g$, for arbitrary $g$.\\
(iii) If $g$ is finite dimensional, $\Lambda^*g$ and $\Lambda^*h$
are isomorphic as Gerstenhaber algebras if and only if $g$ and $h$ are
isomorphic as Lie algebras.

C). A more sophisticated class of Gerstenhaber algebras arise as follows:
Let $M$ ba a manifold (differentiable, complex, algebraic, etc.). Let
$F(M)$ be the commutative algebra of functions (of the appropriate
type --- differentiable, holomorphic, algebraic, etc.) on $M$.
Let $G^*(M)$ be the algebra of polyvector fields on $M$, with
the operations of wedge product and the Schouten bracket, defined
by anology with the bracket in $A^*_n$. Then $G^*(M)$ is
a Gerstenhaber algebra.

The algebras $G^*(M)$ can be thought of in the context of BV
theory: we can regard $G^*(M)$ as the commutative superalgebra of
functions on $\Pi T^*M$, the cotangent bundle of $M$ with
the fibers made into odd supervector spaces. The Gerstenhaber
bracket in $G^*(M)$ is the odd Poisson bracket associated
to the canonical odd symplectic two-form on $\Pi T^*M$.

We should mention an important application of the algebras, $G^*(M)$:
Let $P$ be a bivector field on $M$. We can always construct a bracket
operation on the algebra $F(M)$ by the formula
\be{equation}
\{f,g\}_P=\iota(P)(df\wedge dg)
\end{equation}
where $\iota(P)$ denotes contraction of $P$ against a two-form.
The question is, when does this new bracket $\{,\}_P$ give
rise to a Lie algebra structure on $F(M)$?
\be{pro}\lb{pro1.2}
$\{,\}_P$ satisfies the Jacobi identity if and only if the Schouten
bracket $\{P,P\}=0$.
\end{pro}
Note that when our bracket on functions satisfies Jacobi,
the algebra $F(M)$ becomes what is known in mathematical
physics as a Poisson algebra (see for example \cite{Drin}).

D). Let $N$ be a super-manifold with an odd symplectic structure. Then
the supercommutative algebra of functions $F(N)$ on $N$ has
a structure of a $\bZ/2$-graded, rather than \bZ-graded
Gerstenhaber algebra. We now have the most general context of BV theory,
as described in \cite{W3}. The classical BV master equation reads:
$\{S,S\}=0$ for some even function $S$ in $F(N)$. We see
from Proposition \ref{pro1.2} that the theory of Poisson bracket
structures on function algebras bears a close relation to BV theory.

E). The abstract notion of a Gerstenhaber algebra first arose in
work by M.Gerstenhaber on the Hochschild cohomology of an
associative algebra \cite{Gers1}\cite{Gers2}\cite{GGS}\cite{SS}. In
the very special case that the associative algebra is of the form
$F(M)$ for some smooth manifold $M$, it is a theorem that the
Hochschild cohomology of $F(M)$ is isomorphic as a
Gerstenhaber algebra to the algebra $G^*(M)$ defined in
example C) above. However, in general the Hochschild
cohomology of an associative algebra is known to be
non-isomorphic to an algebra of type $G^*(M)$ (see \cite{GGS}).

{\bf Important Remark:} Just as Poisson algebras are far reaching
generalizations of Poisson algebra of functions on the phase space of
classical mechanics --- i.e. symplectic vector space ---,
Gerstenhaber algebras are far reaching generalizations of
the elementary examples in A) through C) above. There is an overlap
with BV theory, as we explained in the above, but it would be quite
{\em misleading} to identify Gerstenhaber algebra theory with
BV theory.

F). Coboundary Gerstenhaber algebras: in the context of our Theorem
\ref{thm2.2}, we have proved the following (see Lemma \ref{lem2.1}):

{\em On the BRST complex, the following identity holds:}
\[
(-1)^{|u|}\{u,v\}=b_0(u\cdot v)-(b_0u)\cdot v -
(-1)^{|u|} u\cdot(b_0 v)
\]
Thus, the bracket on the BRST complex measures the failure of
the operator $b_0$ to be a derivation of the dot product.
The above statement is the precise analog of an observation
made by Witten in the context of BV theory.
Consider again the algebra $A^*_n$, and define a differential
operator $\Delta$ in $A^*_n$ by the formula:
\be{equation}
\Delta=\sum_i\frac{\partial}{\partial x_i}
\frac{\partial}{\partial x_i^*}
\end{equation}
Then the Schouten bracket in $A^*_n$ measures the failure of $\Delta$
to be a derivation of the dot product in $A^*_n$. We propose the following
abstract definition:
\be{dfn}\lb{def5.3}
Let $G^*$ be a \bZ-graded Gerstenhaber algebra with a linear
operator $\Delta$ of degree -1 such that the following identity holds:
\[
(-1)^{|u|}\{u,v\}=\Delta(u\cdot v)-(\Delta u)\cdot v -
(-1)^{|u|}u\cdot(\Delta v)
\]
\end{dfn}
Then we call the pair $(G^*,\Delta)$ a coboundary Gerstenhaber algebra.

{\bf Remarks:} \\
(i) The motivation for our terminology is simple: the above
equation expresses the fact that (up to certain signs) the bracket,
as a bilinear operator, is the Hochschild coboundary \cite{Gers2}
of the linear operator $\Delta$.\\
(ii) A given Gerstenhaber algebra $G^*$ can be a coboundary
algebra in more than one way: given a $\Delta$ operator as above,
we can simply add any derivation of degree -1 to $\Delta$. Witten
analyzes the case of $G^*(M)$ when $M$ is a manifold, and
observes that any volume form $\Omega$ on $M$
gives rise to a natural $\Delta(\omega)$ operator such
that $(G^*(M),\Delta(\omega))$ is a coboundary algebra. It is
easy to check that for two different volume forms, the corresponding
delta operators differ by a derivation of degree -1.\\
(iii) For any Lie algebra $g$, the Gerstenhaber algebra $\Lambda^*g$
is canonically a coboundary algebra: for the delta operator we may
take the Lie algebra homology differential, $\partial$ (see \cite{Zuck}).\\
(iv) If $M$ is a manifold or variety with singularities, we can
still define the Gerstenhaber algebra $G^*(M)$, but it is
not clear under which circumstances $G^*(M)$ is of
coboundary type, since we cannot simply appeal to the existence of a
volume form.\\
(v) If $N$ is an odd symplectic supermanifold, it is again not
clear whether we have a global delta operator making $F(N)$ into
a coboundary $\bZ/2$-graded Gerstenhaber algebra. BV theory
appears to require the existence of such a delta operator.\\
(vi) If $A$ is an associative algebra, it is not known under
what conditions that Hochschild cohomology of $A$ is a coboundary
Gerstenhaber algebra. The theorem alluded to in E) tells
us that if $A$ is $F(M)$ for a smooth manifold or variety, then
the Hochschild cohomology is in this case a coboundary algebra.

\end{document}